\documentclass[useAMS,usenatbib]{mn2e}

\usepackage{epsfig}
\usepackage{myaasmacros}
\usepackage{amsmath}
\usepackage{amssymb}
\usepackage{gensymb}
\usepackage{graphicx}
\usepackage{deluxetable}
\usepackage{orcidlink}
\usepackage{xcolor}
\usepackage{ulem}

\newcommand{\msun}{M_\odot}

\newcommand{\mbh}{m_{\rm BH}}
\newcommand{\rbh}{\vec{r}_{\rm BH}}
\newcommand{\vbh}{\vec{v}_{\rm BH}}

\newcommand{\rosemary}[1]{\textcolor{black}{ #1}}
\newcommand{\tk}[1]{\textcolor{black}{ #1}}
\newcommand{\hw}[1]{\textcolor{black}{ #1}}
\newcommand{\rs}[1]{\textcolor{black}{ #1}}

\title[Triple SMBHs in Galactic Nuclei]{Analysis of Kozai Cycles in Equal-Mass Hierarchical Triple Supermassive Black Hole Mergers in the Presence of a Stellar Cluster}

\author[Hao et al.]{ 
    Wei Hao$^{1}$\orcidlink{0000-0002-8237-6727},
    M.B.N. Kouwenhoven$^{2}$\orcidlink{0000-0002-1805-0570}, 
    Rainer Spurzem$^{3,4,5}$\orcidlink{0000-0003-2264-7203},  \newauthor
    Pau Amaro Seoane$^{6,7,8,5}$\orcidlink{0000-0003-3993-3249}, 
    Rosemary A. Mardling$^{9}$\orcidlink{0000-0001-7362-3311}, 
    Xiuming Xu$^{2,10}\orcidlink{0000-0001-7643-8621}$ \\
  \\ 
  $^{1}$ Ludwig-Maximilians-Universit\"{a}t M\"{u}nchen, Geschwister-Scholl-Platz 1, D-80539 Munich,  Germany \\ 
  $^{2}$ Department of Physics, School of Mathematics and Physics, Xi'an Jiaotong-Liverpool University, 111 Ren'ai Road, Suzhou Dushu Lake \\ Science and Education Innovation District, Suzhou Industrial Park, Suzhou 215123, P.R.~China \\
  $^{3}$ Astronomisches Rechen-Institut, Zentrum f\"{u}r Astronomie, Univ. of Heidelberg, M\"{o}nchhof-Strasse 12-14, 69120 Heidelberg,\\ Germany \\ 
  $^{4}$ National Astronomical Observatories of China  and Key Laboratory of Computational Astrophysics, Chinese Academy of \\ Sciences, 20A Datun Rd., Chaoyang District, 100012, Beijing, P.R.~China \\
  $^{5}$ Kavli Institute for Astronomy and Astrophysics, Peking University, Yi He Yuan Lu 5, Haidian District, Beijing 100871,\\ P.R.~China\\  
  $^{6}$ Universitat Polit\`{e}cnica de Val\`{e}ncia, Val\`{e}ncia, Spain  \\
  $^{7}$ Max-Planck-Institute for Extraterrestrial Physics, Garching, Germany\\
  $^{8}$ Higgs Centre for Theoretical Physics, Edinburgh, UK\\
  $^{9}$ School of Mathematical Sciences, Monash University, Victoria 3800, Australia\\
  $^{10}$ Department of Physics University of Liverpool, Liverpool L69 3BX, UK\\
}

\begin{document}
\date{Accepted ---. Received ---; in original form ---}
\pagerange{---} \pubyear{2023}
\maketitle
\label{firstpage}


\begin{abstract}
    Supermassive black holes (SMBHs) play an important role in galaxy evolution. Binary and triple SMBHs can form after galaxy mergers. A third SMBH may accelerate the SMBH merging process, possibly through the Kozai mechanism. We use $N$-body simulations to analyze oscillations in the orbital elements of hierarchical triple SMBHs \hw{with surrounding star clusters} in galaxy centers. We find that SMBH triples spend only a small fraction of time in the hierarchical merger phase (i.e., a binary SMBH with a distant third SMBH perturber). Most of the time, the enclosed stellar mass within the orbits of the innermost or the outermost SMBH is comparable to the SMBH masses, indicating that the influence of the surrounding stellar population cannot be ignored. We search for Eccentric Kozai-Lidov (EKL) oscillations \rosemary{for which} (i) the eccentricity of the inner binary and inclination are \rosemary{both oscillate and are anti-phase or in-phase} and (ii) the oscillation period is consistent \rosemary{with EKL timescale. We find that EKL oscillations are} short-lived and rare: the triple SMBH spends around $3\%$ of its time in this phase over the ensemble of simulations, reaching around $8\%$ in the best-case scenario. This suggests that the role of the EKL mechanism in accelerating the SMBH merger process may have been overestimated in previous studies. \rosemary{We follow-up with three-body simulations, using initial conditions extracted from the simulation, and the result can to some extent repeat the observed EKL-like oscillations. This comparison provides clues about why those EKL oscillations with perturbing stars are short-lived. }
\end{abstract}



\section{Introduction}
\subsection{SMBHs and their host galaxies} \label{section:SMBHandhost}


The Milky Way contains an SMBH
at its center \citep{1997MNRAS.291..219G,1998ApJ...509..678G,2017ApJ...837...30G}, and there is convincing observational evidence for the existence of SMBHs residing in the nuclei of all massive galaxies
\citep{1995ARA&A..33..581K,1998Natur.395A..14R}. Various correlations between the central SMBHs and their host galaxy properties have been explored, such as bulge mass and bulge stellar velocity dispersion or dark
matter halo mass, indicating that the formation and evolution of SMBHs
and host galaxies are linked (see, e.g., \citealp{1998AJ....115.2285M, 2000ApJ...539L...9F, 2000MNRAS.311..576K, 2002ApJ...578...90F}; see also \citealp{2013ARA&A..51..511K} for a review). 


Within the framework of a cold dark matter cosmology, dark matter structure and galaxies build up via hierarchical merging \citep[e.g.,][]{1978MNRAS.183..341W, 1991ApJ...379...52W}. As a consequence of the frequent merging of dark matter halos \citep[e.g.,][]{2008MNRAS.386..577F, 2009ApJ...701.2002G} and their corresponding galaxies \citep[e.g.,][]{2008MNRAS.384....2G, 2008MNRAS.383...93B}, the formation of supermassive black hole binaries (SMBHBs) might be a natural consequence of mergers of two galaxies with pre-existing black holes \citep{1980Natur.287..307B}. In particular massive early-type galaxies, hosting the most massive black holes in the Universe might even have experienced a significant number of mergers with other more or less massive galaxies during the late phases of their assembly \citep[e.g.,][]{2006ApJ...640..241B, 2006MNRAS.366..499D, 2007ApJ...654..858B, 2006ApJ...652..270B,2008ApJ...688..770F, 2010ApJ...709.1018V, 2010ApJ...715..202H, 2013MNRAS.428.3121M,2013ApJ...770...57B}. These galaxy merger events provide a natural explanation for the structural evolution of the massive early-type galaxy population \citep[e.g.,][]{2006MNRAS.369.1081B, 2006ApJ...636L..81N, 2009ApJ...697.1290B, 2009ApJ...699L.178N, 2009ApJ...695..101D, 2009ApJ...706L.120V, 2009ApJ...697.1369B, 2010ApJ...719..844R, 2010MNRAS.401.1099H, 2012ApJ...744...63O, 2013MNRAS.429.2924H, 2013arXiv1311.0284N}; see \citealp{2017ARA&A..55...59N} for a review. 


Mergers between massive galaxies occurred regularly in cosmic history. It is plausible to assume that their black holes merged as well. However, the efficiency of the SMBH merger process is actively debated \citep{2003ApJ...596..860M, 2013ApJ...773..100K, 2014ApJ...785..163V, 2017MNRAS.464.2301G, 2022A&A...665A..86B}. Galaxy mergers can be broadly divided into minor mergers and major mergers, depending on the mass ratio of the galaxies. In a minor merger (mass ratio smaller than $1:10$), the mass of the satellite galaxy can be so small that dynamical friction becomes inefficient \citep{2009ApJ...707L.184J, 2011ApJ...729...85C, 2012ApJ...756...30K, 2014MNRAS.439..474V} and the merger timescale might be longer than the age of the Universe \citep{2001ApJ...563...34M, 2001ApJ...549..192P, 2002MNRAS.331..935Y}. For major merger events (mass-ratio greater than $1:10$), gas-rich galaxies merge much more efficiently than gas-poor galaxies, their merger timescale can be relatively short, and their central black holes can sink rapidly to the center of the remnant
\citep{1997NewA....2..533Q, 2001ApJ...563...34M, 2002MNRAS.331..935Y,2008MNRAS.383...93B}. For massive and gas-poor early-type galaxy mergers, the merger timescale can be comparable to the age of the Universe. Their long-lasting merger process can be divided into three evolutionary phases \citep{1980Natur.287..307B}. In the first phase, the infalling galaxy centers (and their SMBHs) are slowed down through the dynamical friction exerted by surrounding stars and dark matter (e.g. \citealp{1988ApJ...331..699B}), and the SMBHs sink rapidly to the center as a bound binary \citep{1996NewA....1...35Q}. Thereafter, slingshot ejection of surrounding stars on centrophilic orbits becomes the dominant process that takes energy and angular momentum away from the SMBHB and thus leads to the shrinking of the separation between the SMBHBs \citep{1962rdgr.book..367P, 1973ApJ...183..657B, 1984MNRAS.211..933F, 2001ApJ...563...34M, 2002ApJ...567..817H, 2005LRR.....8....8M, 2006ApJ...642L..21B,2007ApJ...666L..89L,2008ApJ...676...54C}. Finally, after the SMBHB reaches a separation close enough (e.g.,  $\sim$  0.1~pc for $10^8 \msun$ SMBHB) to trigger significant gravitational wave (GW) emission, the binary quickly loses binding energy and may finally coalesce \citep{1964PhRv..136.1224P,2000ApJ...528L..17P,2011ASL.....4..181C}. 
Low-mass black holes grow mostly by gas accretion, but towards higher masses, their host galaxies become more gas-poor, so the direct merging of SMBHs becomes more important. It might even dominate the mass growth with more than $50\%$ for SMBHs beyond 
$10^9 \msun$ \citep{2007MNRAS.382.1394M, 2013ApJ...768...29V, 2015ApJ...799..178K}; see, however, \cite{2004ApJ...615..130M} showing that the overall contribution of merging is small. Many black holes are in binary systems, in particular at high redshift \citep{2003ApJ...582..559V}. The merging of SMBHs might even support the formation and explain the tightness of relations between black hole mass and galaxy mass properties \citep{2007MNRAS.382.1394M, 2010MNRAS.407.1016H,2011ApJ...734...92J}.

Numerical studies of spherically symmetric galaxies suggested slingshot ejection of stars on centrophilic orbits is not sufficient to bring an SMBHB close enough to trigger GW emission within a cosmic time. This so-called 'final parsec problem' \citep{2003ApJ...596..860M} is caused by the depletion of the 'loss cone' \citep{1976MNRAS.176..633F,2001MNRAS.327..995A,2001ApJ...563...34M}, a region in stellar phase space available for slingshot ejection. Further studies
suggest that triaxial or axisymmetric galaxies harbor more stars on centrophilic orbits, reducing the SMBH merging time (e.g., \citealp{2002MNRAS.331..935Y, 2004ApJ...606..788M, 2006ApJ...642L..21B, 2011ApJ...732...89K, 2013ApJ...773..100K}; see, however, \citealp{2014ApJ...785..163V, 2015ApJ...810...49V}). However, this long period ($\sim 0.1-1$~Gyr for the coalescence of SMBHBs with mass $\sim 10^8 \msun$; \citealp{2012ApJ...749..147K}) would give a chance to the formation of a secondary merger, when the merger timescale is much larger than the interval between two mergers. Since most massive galaxies should have experienced multiple mergers during their lifetime, triple or higher-order SMBH systems may form. When a third galaxy falls towards the ongoing merger of SMBHB, a hierarchical triple system can form \citep{2002ApJ...578..775B, 2007MNRAS.377..957H}. If we consider the central inner SMBHB as a subsystem with respect to the third SMBH on a wider orbit, we can
describe the system as consisting of an inner binary and an outer binary system, in a   hierarchical triple system. 


\subsection{Nuclear star clusters} \label{section:NSCs}

\rs{Most if not all galaxies host nuclear star
clusters (NSCs) in their center, typically with masses in the range $10^8 \sim 10^{10} \msun$ \citep[cf. for this and the following the review of][]{2020A&ARv..28....4N}. These extremely dense stellar systems are highly luminous sources that stand out above their surroundings. NSCs and massive black holes are found to exist together \citep{2003ApJ...588L..13F, 2009MNRAS.397.2148G}, the most well-known example for this is our Milky Way, which is the host of an inactive SMBH with $\sim 4 \cdot 10^6 \msun$ that resides in a
NSC with $ \sim 10^7 \msun$ \citep{2010RvMP...82.3121G}.}
\rs{NSC reside within the central few tens of parsecs of a galaxy;
the NSC of our own Galaxy is very well studied, its stellar distribution}
\hw{ \citep{2003ApJ...586L.127G,2003Natur.425..934G,2017ApJ...837...30G}, as well as its dynamical properties \citep{2009ApJ...690.1463L,2009ApJ...697.1741B,2015A&A...584A...2F}. NSCs have been discovered in many other galaxies throughout the universe \citep{2002AJ....123.1389B,2004AJ....127..105B,2004ARA&A..42..603K,2006ApJS..165...57C}, see \citealp{2020A&ARv..28....4N} for a review. }

\hw{The formation of nuclear star clusters is still an area of active research, and there is no universally accepted model for their origin. However, several mechanisms have been proposed to explain their formation. Some NSCs are believed to form in situ, which means that they originate from the material inside their host galaxies \citep{2011ApJ...729...35A,2021A&A...650A.137F}.}
\rs{Another model suggests that the migration of globular clusters toward their galactic centers leads to the formation of an NSC and may also support the growth of an SMBH by intermediate-mass black holes falling in with the clusters \citep{2014A&A...570A...2F,2023A&A...674A.148N}. \citealp{2017MNRAS.467.4180S} developed an analytical model of the growth of massive black holes in a NSC.}
\hw{Correlations between NSCs and their central SMBHs suggest}
\rs{that their evolution is coupled \citep{2012AdAst2012E..15N,2015ApJ...803...81N}.}


\subsection{SMBH triples} \label{section:SMBHT}


The influence brought by a third galaxy on accelerating the merger process of a progenitor SMBHB system is multi-faceted. As the SMBH from the infalling galaxy slowly sinks towards the center, it perturbs the progenitor SMBHB repeatedly, exciting the eccentricity of the binary. This may accelerate the merging process by increasing the release of gravitational radiation \citep{1994ApJ...436..607M}. A side effect of the eccentricity growth of the SMBHB is that the cross-section for the slingshot mechanism is slightly altered as well, and therefore more stars are ejected due to the change of their loss cone. Besides the influence on the SMBHB, the gravitational potential of galaxies is changed by the infall of the third galaxy. This produces additional stars on boxy orbits, which refill the loss cone of the progenitor SMBHBs \citep{2006ApJ...636L..81N}. Also, the presence of additional stars brought in by the third galaxy provides more stars with centrophilic orbits, which are ejected by the progenitor SMBHB, which accelerates the merger process \citep{1988ApJ...334...95R}. Finally, the presence of a tertiary SMBH could also impact the merger process of the binary \citep[Xu et al., in prep]{2002ApJ...578..775B}. As a result, the merger timescale of the progenitor SMBHB can be dramatically reduced. Alternatively, instead of forming interacting triple SMBH systems in a galaxy center, a free-floating SMBH may also emerge when one of the SMBHs in the triple system is ejected out by slingshot interaction \citep[typically the least massive body is ejected; see][]{1994AAS...184.3305V}. A merger with GW emission of two components of the triplet could happen before the final ejection of one component, depending on the properties of the host galaxy centers and the properties of the SMBHs \citep[e.g., mass ratio, separations, and velocities; see ][]{2006ApJ...651.1059I,2007MNRAS.377..957H,2008arXiv0801.0859I,2010MNRAS.402.2308A,2018MNRAS.477.3910B}. 


The process described above is complicated, and one natural aspect of simplifying this process is to ignore the other parts of galaxies and focus on the hierarchical SMBH triple only. Even the systems with hierarchical SMBH triple only are complicated, since general relativity (GR) must be considered beside the gravitational triple system. So here we put aside the GR effect first and consider a simple case with hierarchical triple stellar systems instead. In such systems, while the outer binary and inner binary torque each other and exchange angular momentum, their eccentricities experience periodic oscillations over secular timescales under certain conditions. These oscillations usually occur on a secular timescale that is much longer than their orbital periods \citep{2000ApJ...535..385F}. For non-coplanar orbits, the inclination angles also have corresponding oscillations. According to canonical perturbation theory, no oscillation in the semi-major axis is predicted, since the energy exchange averages out during the secular evolution \citep{1975MNRAS.173..729H}. However, \cite{1979A&A....77..145M} find that for triple stellar systems, the net effect of tidal dissipation together with eccentricity perturbations may lead to significant decay of the eccentricity and the semi-major axis of the inner binary. 


Since SMBHs are massive compact objects, general-relativistic precession of the inner orbit should be taken into account. When the precession period from general relativity is comparable to that of the Newtonian perturbations, it could either enhance the periodic oscillation generated by the orbital perturbations \citep{1984A&A...141..232S} or it could break down the oscillation \citep{1997Natur.386..254H}. Thus, to further simplify the problem, if we ignore general relativity effects, tidal disruption, and the spin of SMBHs, considering only the Newtonian gravitational interactions between the hierarchical triple SMBHs, then the merger process can be simplified as an isolated restricted three-body problem. 


\subsection{The Kozai-Lidov mechanism} \label{section:KLM}


When the inner and outer orbital planes are mutually inclined, this system can be considered as a triple system following the Eccentric Kozai-Lidov (EKL) mechanism, which refers to the Standard Kozai-Lidov (SKL) mechanism releasing the test-particle quadrupole (TPQ) approximation and the circular outer orbit constraints \citep{2000ApJ...535..385F, 2011PhRvL.107r1101K, 2011ApJ...742...94L,2011Natur.473..187N,2013MNRAS.431.2155N}. Von Zeipel, Kozai and Lidov discovered independently that for a hierarchical triple system, with one component of the inner binary harboring negligible mass and the outer binary on a circular orbit, if the orbital plane of the inner binary and outer binary are mutually inclined ($39.2^\circ < i < 140.8^\circ$), then the orbital eccentricity and inclination of the test particle orbit will oscillate with an opposite phase, the so-called Standard Kozai-Lidov mechanism \citep{1910AN....183..345V,1962AJ.....67..591K, 1962P&SS....9..719L,2019MEEP....7....1I}. Furthermore, \cite{1968AJ.....73..508H} studied the general mass ratio for hierarchical triples and obtained a similar quadrupole approximation. \cite{1984A&A...141..232S} derived octupole level equations in the limit of low eccentricities and inclinations and showed that the octupole approximation plays an important role than the quadrupole approximation in this regime. 


More detailed studies on the EKL mechanism have been carried out recently and flipping of the inner orbit inclination orientation from prograde to retrograde has been found (see \citealp{2016ARA&A..54..441N} for a brief review and \citealp{2017ASSL..441.....S} for a longer one). \cite{2005MNRAS.358.1361I} discussed the changes in angular momentum within an orbit while studying the EKL cycles in tidal disruption of stars near unequal-mass SMBH binaries.\hw{Various definitions of the hierarchy have been further explored for which the averaging approximation is valid \citep{2011ApJ...742...94L,2014MNRAS.439.1079A,2011PhRvL.107r1101K,2014MNRAS.438..573B}.  \cite{2021AJ....161...48B} discussed the limit within which the secular approximation is reliable for a mildly hierarchical triple when the outer perturber is in a circular orbit. \cite{2013MNRAS.430.2262H,2015ApJ...799...27P,2018MNRAS.481.4907G,2018MNRAS.476..830M} and \cite{2023ApJ...952..103Z} also explored the stability of the hierarchical triple system and applicability of the double-averaged procedure under different schemes.} \cite{2014MNRAS.439.1079A} reported failure in simulating the long-term evolution of EKL oscillations using the double-averaging approximation. Similarly, \cite{2016MNRAS.458.3060L} find rapid eccentricity oscillations in addition to secular oscillations, while using direct $N$-body simulations instead of secular approximations, and thereafter developed corrected equations for the double-averaging approximation, and after the correction it agrees with the $N$-body simulation result. 

For the hierarchical triple SMBH systems with comparable masses and random configurations of their orbits, neither the TPQ approximation nor the circular outer binary orbit constraint is satisfied, but still similar oscillations can be observed \citep{2002ApJ...578..775B}. 


Three-body systems with certain initial conditions can be chaotic systems \citep[e.g.][]{2014ApJ...791...86L,2021AJ....161...48B,2020MNRAS.499.1682H}. Even the tiny differences between secular approximation and direct $N$-body method on simulating isolated hierarchical triples can lead to different evolutionary results, as well as their oscillation patterns. This implies that it is difficult to study the dynamical evolution of isolated SMBH triples in galaxy centers with surrounding nuclear star clusters or even more extended galactic bulges. However, as what \cite{1976CeMec..13..471L} called a happy coincidence, this secular problem is integrable at the quadrupolar approximation\citep{2010MNRAS.401.1189F}. Fortunately, statistical studies are proven to be effective \citep{1999MNRAS.310..811H, 2019Natur.576..406S, 2020MNRAS.497.3694M, 2020MNRAS.493.3932B}. In this regime, would those EKL oscillations still be observed as expected in previous studies, or would they become short-lived even no longer exist? \cite{2010MNRAS.402.2308A} simulated SMBH triplets in galactic centers with a focus on the cumulative eccentricity distribution during the merger of the inner binary and their gravitational wave radiation signal prediction. Although EKL is believed to play a pivotal role during the evolution of hierarchical triple SMBHs in galaxy centers, it is not clear to what extent this process affects the merger probabilities of the inner SMBHBs. Detailed statistics on the properties of these oscillations, for example, how their timescale is comparable to the EKL timescale, how often these oscillations occur, and how long they last, are required. 


As in the SKL mechanism, oscillations driven by the EKL mechanism can also repeatedly excite the eccentricity of the inner binary, which may lead to a fast coalescence of the merging SMBHB since gravitational wave emission efficiency is strongly related to the eccentricity of the inner binary \citep{1964PhRv..136.1224P}. Furthermore, \cite{2020MNRAS.495.2321Z} used a revised formula to improve Peter\rq{}s estimation, showing there is a 1-10 times deviation while using Peter\rq{}s formula. 
The effect of the EKL mechanism on accelerating the merger of the inner binaries in isolated hierarchical triple SMBH systems has been studied by \cite{2002ApJ...578..775B} using the orbital averaging method for three-body integrations, showing that the presence of the third SMBH can shorten the merger timescale of inner SMBH binary by an order of magnitude in more than a half of all cases. 


Few active galactic nuclei (AGN) pairs, candidate SMBHBs, with separation below $\sim 1$~kpc have been identified so far; see, for example, \cite{2016arXiv160606568K} for a brief review. However, recent observations of triple massive black hole systems show that triple SMBH systems appear to be more common than previously believed \citep{2014Natur.511...57D, 2020A&A...633A..79K, 2021ApJ...907...72F}. The final coalescence of an SMBHB is the most energetic event as well as the most powerful GW emitter in the Universe. The detection of GWs from a coalescing SMBHB would be important since it would provide robust evidence of the existence of black holes and examine Einstein's relativity equations in the strong-field limit. It is expected to be measured in the future, for example, by LISA \citep{2014AAS...22324813H, 2017arXiv170200786A, 2022arXiv220306016A,2023LRR....26....2A}, Pulsar Timing Arrays \citep{2005ASPC..328..399J, 2016MNRAS.458.1267V}, and SKA \citep{2015aska.confE..37J, 2020PASA...37....7S}. For SMBHs with a mass over $10^8 \msun$, their GW wavelength will be detectable only by PTAs. 


This study aims to analyze the data from the $N$-body simulation on hierarchical triple SMBHs residing in the center of gas-poor elliptical galaxies, with a focus on the oscillation verification of known mechanisms, such as near-Kozai oscillation and three-body oscillation generated by energy or angular momentum exchanges between the orbits, as well as other oscillation patterns that are not observed in pure three-body interactions but found in those SMBH triples residing in galactic centers. 
Although the Kozai mechanism is believed to be an important process in many processes related to three-body interaction at SMBH vicinity in previous studies, it is currently difficult to observe directly. This work is aimed at analyzing simulated data as a prediction before we can get a direct observation, and get a better understanding of related physical processes. Due to the complications discussed above and in order to study the impact of the surrounding star or star clusters on the SMBHs, we focus mainly on the simplified situation of equal-mass triple SMBH systems embedded in a stellar environment. Thus, we do not include the GR effect and tidal effects in the simulation but we compute the timescale instead. We use period analysis to extract the oscillation information from the dynamical evolution of the detailed $N$-body simulations of these hierarchical triple SMBHs in galactic centers, especially oscillations related to the semi-major axis and eccentricity of the inner binary. Then we analyze the relatively long-term periodic behavior with respect to their orbital motions and summarise the patterns of the oscillation and their frequencies. 


This paper is organized as follows. In Section~\ref{section:method} we introduce our simulation and initial parameter selection. In Section~\ref{section:analysis}, different points of aspects on analyzing the evolution process are explored, as well as a detailed discussion on both the properties of different types of oscillations and their influence on the merger process. Isolated comparison three-body simulations with the same initial condition are presented in Section~\ref{section:comparison3body}. Finally, we summarise and discuss our conclusions in Section~\ref{section:conclusions}.


\section{Method and initial conditions} \label{section:method}

\begin{table}
 \centering
  \begin{tabular}{l rrr rrr}
        \hline
                & A		    & B			& C			& D			& E			& F			\\
    \hline 
    $x_1$       & 0.0		& 	0.0		&	0.0		&	0.0		&	0.0		&	300	\\
    $y_1$       & 200.0	    &  	400.0	&	200.0	&	400.0	&	100.0	&	5.0		\\
    $z_1$		& 0.0		&	0.0 	&	0.0 	&	0.0 	&	0.0 	&	0.0 	\\      
    $x_2$		& --173.2 	& 	--88.2	&	--173.2	&	--88.2	&	--88.2	&	--173.2 \\
    $y_2$		& --100.0	&	--50.0	&	--100.0	&	--50.0	&	--50.0	&	100.0	\\      	
    $z_2$		& 0.0		&	0.0		&	250.0 	&	250.0	&	0.0 	&	250.0 	\\
    $x_3$		& 173.2 	& 	88.2	&	173.2	&	88.2	&	88.2	&	173.2  \\
    $y_3$		& --100.0	&	--50.0	&	--100.0	&	--50.0	&	--50.0	&	100.0 	\\
    $z_3$		& 0.0		& 	0.0 	&	0.0 	&	0.0 	&	0.0 	&	0.0 	\\   
    \hline
    $V_{x1}$ 	& --21.92	&  	--21.92	&	--21.92	&--205.25	&	21.92	& 	--160.57	\\
    $V_{y1}$  	& 0.0 		&  	0.0		&	0.0		&	0.0		&	0.0		&	102.63	\\
    $V_{z1}$  	& 0.0    	&  	0.0		&	0.0		&	0.0		&	0.0		&	0.0	\\ 
    $V_{x2}$ 	& 10.96	    & 	10.96	&	10.96	&	10.96	&	10.96	&	10.96	\\
    $V_{y2}$	& --18.99	& 	--18.99	&	--18.99	&	--18.99	&	--18.99	&	--18.99	\\
    $V_{z2}$	& 0.0 		& 	0.0		&	0.0		&	0.0		&	0.0		&	0.0	\\
    $V_{x3}$ 	& 10.96	    & 	10.96	&	10.96	&	19.76	&	10.96	&	10.96	\\
    $V_{y3}$	& 18.99  	& 	18.99	&	18.99	&	18.99	&	18.99	&	18.99	\\
    $V_{z3}$	& 0.0		&	0.0		&	0.0		&	0.0		&	0.0		&	0.0	\\
    
    \hline
  
  \end{tabular}
  \caption{Initial positions (in pc) and velocities (in km/s) for the central triple SMBHs in 6 simulations (models A--F). Subscript 1 denotes the position or velocity for the first SMBH, subscript 2 for the second, and subscript 3 for the third. The initial mass of each SMBH is $10^8 \msun$, and the total mass is $10^{10} \msun$ for all 64\,000 particles. \label{table:initialBHs} }
\end{table}

\begin{figure}
	\centering   
  \includegraphics[width=0.5\textwidth,height=!]{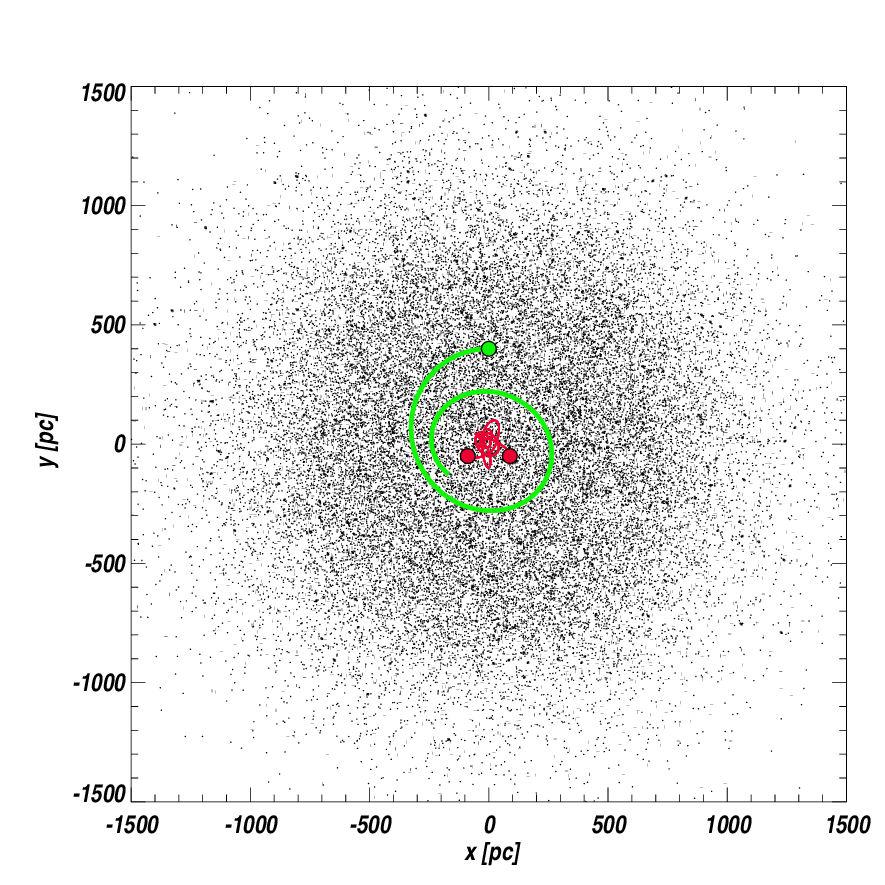}
  \caption{Example of the initial condition of the hierarchical SMBH triple. The two red dots represent the inner binary, and the green dot represents the third outer black hole. The curves show the trajectories of the three SMBHs at the beginning of the simulation, red for the inner binary and green for the third SMBH. Black dots indicate the stellar population (each with a mass of $1.53 \times 10^5 \msun$ if we scale each SMBH mass to $10^8 \msun$) that represents the galactic centre. 
     \label{figure:initialcon} }
\end{figure} 


We carry out direct-summation $N$-body simulations to determine the dynamical evolution of the SMBH triples. The $N$-body simulations are initialized with the total particle number of $N=64,000$, including three SMBHs. Each SMBH has a mass of 1\% of the total mass. The $N_\star=63\,997$ surrounding equal-mass bodies that make up the remaining 97\% of the total mass. The \citep{1911MNRAS..71..460P} model with an $n=5$ polytrope is adapted to generate the initial positions and velocities of the stellar system. We adopt $N$-body units, which are scale-free \citep{1986LNP...267..233H}. For our proposed scheme, setting the mass of each SMBH as $10^8 \msun$ and the length scale fitting the center region of massive elliptical galaxies ($500$~pc for 1 $N$-body unit), then the total mass of the galactic nucleus is $10^{10} \msun$, and the $N$-body time unit is $1.6938$~Myr. All our particles are non-evolving point masses. 


\hw{The three SMBHs are initialized as binary SMBHs plus a third SMBH (see Figure~\ref{figure:initialcon}), in which the separation of two SMBHs is about one order of magnitude smaller than the distance to the third SMBH (see Table~\ref{table:initialBHs}) \citep{2010MNRAS.402.2308A}. The third SMBH is initially loosely bound or even unbound (e.g. in Model D) to the inner SMBH binary but note that the out-most SMBH will later pulled back by the stellar cluster due to dynamical friction instead of escape.} The three black holes are initially placed within the Plummer radius, with velocities below their escape velocities. We distinguish inner and outer binaries by comparing the relative distances of the three SMBHs. The initial positions and initial velocities of the SMBH triple in each simulation are well spread in parameter space in order to represent different initial conditions. 

\rs{Plummer's model is not a suitable initial density distribution for NSCs \citep[cf. e.g. ][]{1976MNRAS.176..633F,1976ApJ...209..214B,2010RvMP...82.3121G}; it is chosen here for simplicity and since within a fraction of a relaxation time a central density cusp around the innermost SMBHs does form \citep[][for a single SMBH]{2004ApJ...613L.109P}; this method has been used also in our previous work \citep[e.g.][]{2019MNRAS.484.3279P}. In case of our system of triple SMBH the cusp forms around the inner pair of SMBH. The stellar density distribution around binary and triple SMBH after a galactic merger is complex, may exhibit rotation and bars and is currently beyond the scope of our work.
}

Since the merger time scale of dark matter halos and bulges is much shorter than the typical time interval between two consecutive mergers, the merging of these components except SMBHB should be accomplished before the third SMBH approaches the central SMBHB. This also provides a pathway for the inner binary to dynamically evolve, resulting in a low-eccentricity orbit due to the interaction with surrounding matter. Thus, the initial eccentricity of the inner binary should be low, and the initial separation should be close to the hardening radius \citep{1996NewA....1...35Q}.  Studies on the coalescence of SMBHBs have shown that the separation of the binary would decrease slowly after they reach this third phase, relying on sling-shot ejections of surrounding stars, typically, at several parsec distances for $10^6 \msun$ SMBHBs, or several tens parsec for $10^8 \msun$ SMBHBs, respectively. 


In this scheme, the typical initial semi-major axis of the inner binary is of order ten parsecs, while the outer binary is approximately ten times wider.  The simulation is terminated once the inner binary reaches a separation of sub-parsec scale, which is the typical distance for a $10^8 \msun$ SMBHB with a highly eccentric orbit to trigger an immediate merger (within 1~Myr) due to GW emission \citep{2012ApJ...749..147K}. In other words, the total simulation time depends on the efficiency of the merger process. For simplicity, we do not include the general relativistic effect. Instead, we compute the timescale of the GW emission to have a sense of how strongly it may affect the actual motion of SMBHs. 


Our simulations are carried out using the direct $N$-body code \texttt{NBODY6++}, the parallel version of Aarseth\rq{}s \texttt{NBODY6} \citep{1999JCoAM.109..407S}. The code does not use a softening to the gravitational force in order to attain high accuracy. Instead, the Kustaanheimo and Stiefel (KS) regularisation has been employed to close binaries \citep{1965JRAM..218..204P}. By carrying out an ensemble of 1000 sets of three-body simulations with widespread initial conditions, \cite{2010MNRAS.402.2308A} drew the conclusion that information about initial conditions will be rapidly erased by chaotic interactions between SMBH triples, leading to a result of similar statistical characteristics for different initial inclination distributions. The influence of initial conditions is erased shortly after the simulations started. In spite of this, different initial conditions with different energy levels would have different evolution with higher or lower fraction time in binding status, as well as different oscillation types and evolutionary timescales before the inner binary starts to merge. To ensure the diversity of our investigation, the initial orbital parameters and the alignment of three black holes are well spread in the parameter space (Table~\ref{table:initialBHs}). 

In order to study the effect brought by the surrounding star clusters, we take snapshots of the SMBH triples and extract their position and velocity information, then run another set of simulations with the SMBH triple only for comparison. This comparison could give us an idea of how the surrounding stars could affect the central SMBH triples. In the target region that we selected to do the comparison simulation, we obtain snapshots at an interval of 0.1 $N$-body time units. Detailed method descriptions are shown in \S~\ref{section:comparison3body}. 


\section{Data analysis and results} \label{section:analysis}

\subsection{General properties of orbital evolution} \label{section:orbitalevolution}

\begin{figure*}
  \vspace*{50pt}
  \includegraphics[width=0.95\textwidth,height=!]{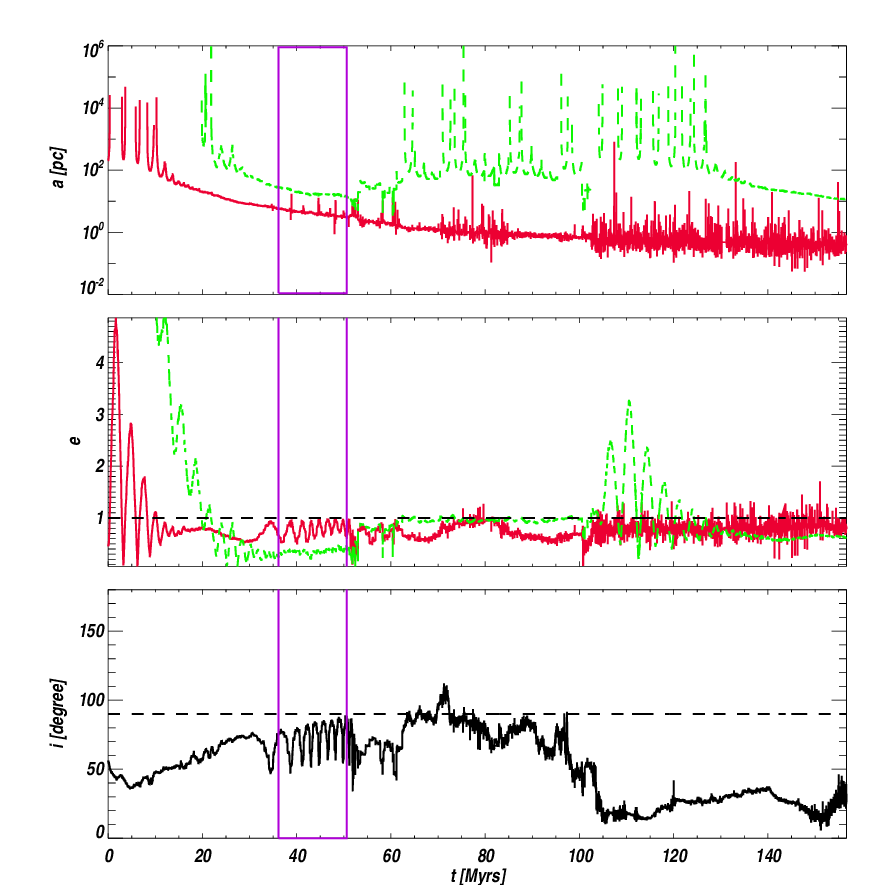}
  \caption{Evolution of the semi-major axis ({\em upper panel}), eccentricity ({\em middle panel}), and relative inclination between the inner and outer orbits ({\em bottom panel}), in simulation model D. The red curve represents orbital parameters of the inner binary and the green curve represents the outer binary. The discontinuities in the semi-major indicate times in which the three-body system is unbound. \hw{The magenta box highlights a selected time span during which clear oscillations are observed.}
     \label{figure:fullevolution} }
\end{figure*} 

\begin{figure*}
 	\vspace*{50pt}  
  \includegraphics[width=1.0\textwidth,height=!]{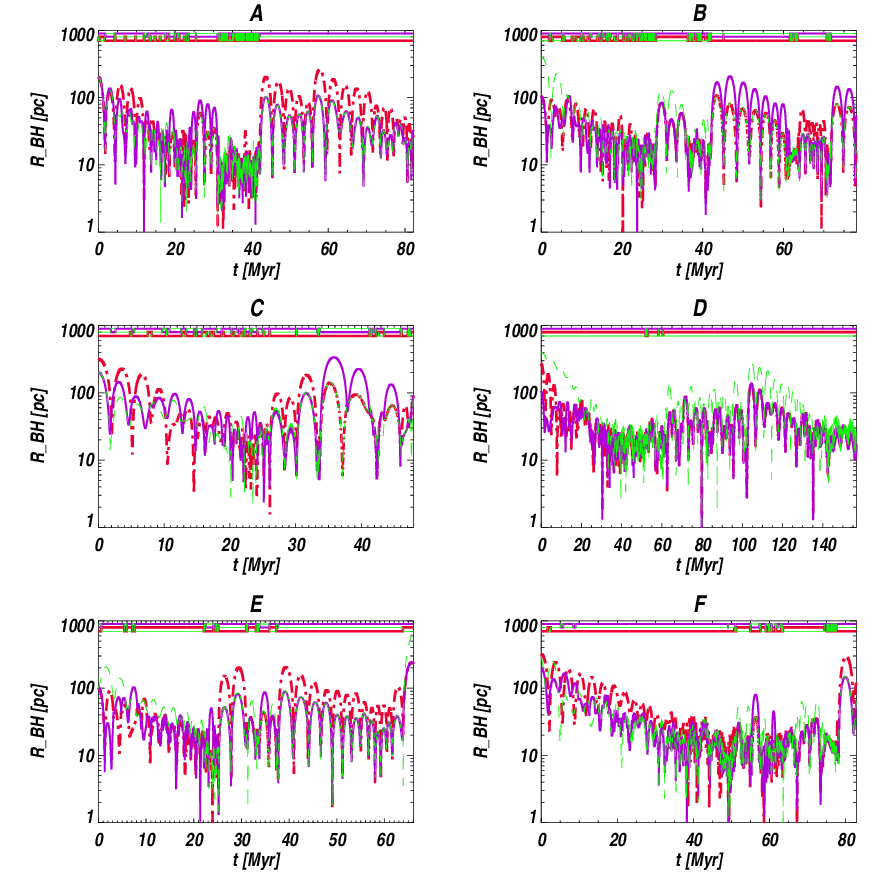}
  \caption{Distance between the SMBHs and the enter-of-mass of the galactic nucleus, as a function of time. The green, red, and magenta curves represent the first, second, and third SMBH, respectively. Note that when two of the SMBH form a close binary, their distances are nearly identical and the curves overlap. The horizontal curves at the top of each panel show the switches in the order of the hierarchical triple due to the exchanges of the members of the inner and outer binary, since here (and only for this figure) we define the inner binary as the pair with the smallest separation. }
     \label{figure:sixinone} 
 \end{figure*} 

\begin{figure*}
 	\vspace*{50pt}  
   	\includegraphics[width=\textwidth,height=!]{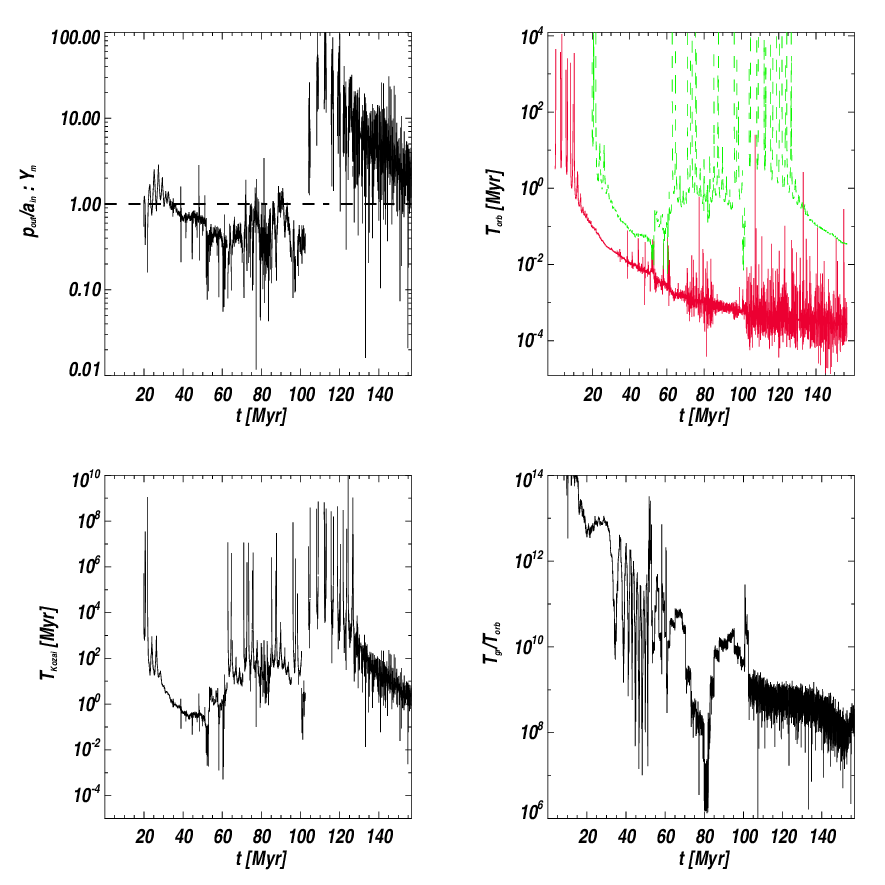}
   	\caption{Evolution of several timescale parameters \rosemary{for model D}. {\em Upper-left panel:} Mardling \& Aarseth criterion (with inclination modification), the system is stable if the value is larger than unity. {\em Upper-right panel:} orbital time evolution for the inner (red) and outer (green) binary. {\em Lower-left panel:} critical period of the Kozai oscillation computed from the orbital elements of three black holes. {\em Lower-right panel:} the ratio between gravitational wave emission time scale and orbital period of the inner binary.
   	\label{figure:keyparameters} }
\end{figure*}


To analyze the evolution of the triple SMBHs we adopt Jacobi coordinates. The masses of the inner binary are $m_1$ and $m_2$, respectively, the distant third black hole has a mass of $m_3$, the semi-major axis of the inner binary is $a_{\rm in}$, with eccentricity $e_{\rm in}$, and the semi-major axis and eccentricity of the outer binary (i.e., the system containing the third black hole and the center-of-mass of the inner binary) are $a_{\rm out}$ and $e_{\rm out}$. The mutual inclination angle between the orbital plane of the outer binary and the orbital plane of the inner binary is $i$. We adopt this notation even when the hierarchy status of the system switches, i.e., when one component of the inner binary is exchanged with the outer SMBH, or even a component is ejected. With the center-of-mass of the inner binary $M_{\rm in}$, the gravitational constant $G$, the angular momentum $\vec{L}_{\rm in}$, and the total energy of the inner binary $E_{\rm in}$ we compute the semi-major axis 
\begin{equation}
	a_{\rm in} = - \frac{G m_1 m_2}{2E_{\rm in}} 
	\quad,  
\end{equation}
and the eccentricity
\begin{equation}
	e_{\rm in} = \sqrt{1 + \frac{2E_{in} |\vec{L}_{\rm in}|^2}{M_{\rm in} (G m_1 m_2)^2}} 
	\quad.  
\end{equation}
Similarly, we have for the outer binary  
\begin{equation}
	a_{\rm out} = - \frac{G M_{\rm in} m_3}{2E_{\rm out}} 
	\quad,   
\end{equation}
and
\begin{equation}
	e_{\rm out} = \sqrt{1 + \frac{2E_{\rm out} |\vec{L}_{\rm out}|^2}{M_{\rm out} (G M_{\rm in} m_3)^2}} 
            \quad, 
\end{equation}
where $M_{\rm out}$, $\vec{L}_{\rm out}$, and $E_{\rm out}$, are the center-of-mass, the relative angular momentum of the outer binary, the total energy, and the $z$-component of relative angular momentum of the outer binary, respectively. The mutual inclination angle between the orbital planes of the inner binary and the outer binary is 
\begin{equation}
	i = \arccos{ \frac{\vec{L}_{\rm in}{\cdot}\vec{L}_{\rm out}}{|\vec{L}_{\rm in}|\ |\vec{L}_{\rm out}|}} 
	\quad.  
\end{equation}
%


Figure~\ref{figure:fullevolution} shows the evolution of the semi-major axis, eccentricity, and the relative inclination of the orbital planes of the inner and outer binary, for simulation~D (see Table~\ref{table:initialBHs}). 


At the start of simulation~D, the inner binary has a semi-major axis of order 1~kpc, while the outer SMBH is unbound to the inner binary. As a result of interactions and exchanges, the semi-major evolution track is discontinuous. Note that even for the inner binary, the data of semi-major axis evolution are sometimes discontinuous because the SMBHs are sometimes unbound to the binary system (but still bound to the cluster), and the semi-major axis becomes temporally negative. Although these binaries appear temporally unbound, they do not escape from the system in most cases, since dynamical friction and frequent encounters with surrounding stars would pull them back towards the cluster center (see their distance to the center in panel~D of Figure~\ref{figure:sixinone}). During this journey (trying to escape and pulled back), the escaping SMBH loses part of its kinetic energy and the triple system becomes more stable. The net effect of these processes leads to the inner binary semi-major axis shrinking rapidly to roughly an order of hundred parsecs within the first 20~Myr, while the semi-major axis of the outer binary transforms its status from unbound to stably bound. When the inner binary or the outer binary is unbound, or the outer binary is too wide to effectively interact with the inner binary, we refer to this as the unstable phase. 


In the middle stage, the third black hole approaches the inner binary, and violent encounters start to play an important role. Close encounters then lead to chaotic motions and frequent exchanges of inner and outer components, as well as kicking out of one component temporally, therefore resulting in intermittent unbound of the binaries, especially the outer binary. Occasionally, after some close encounters between the three SMBHs (e.g., at $t \approx 50$~Myr in Figure~\ref{figure:fullevolution}), the inner semi-major axis temporarily becomes close to that of the outer semi-major axis, which means components of inner and outer binary occasionally experience exchanges. Both of these unbound states and exchanges can be put down to the disturbance exerted by the galaxy center environment and the instability of the three-body system itself. In order to determine how often these exchanges occur, we redefine the inner and outer binary by comparing their mutual distances and show these switches with horizontal lines at the top of each panel in Figure~\ref{figure:sixinone}. When these exchanges occur, the triple SMBHs are all relatively close to each other, and the three-body system is temporally unstable. We thus refer to this as the chaotic interacting phase. 


In the final stage, both the inner binary and the outer binary become sufficiently hard, and the three SMBHs mostly retain a hierarchical triple structure. We refer to this as the hierarchical merger phase. During this stage, the inner binary approach is closer to each other efficiently under the repeated perturbations of the third SMBH and surrounding stellar components \hw{until the separation between the inner SMBH binary reaches the sub-parsec scale and merges due to GW emission. }


\rs{Close interactions between stellar objects and central SMBHs lead to a number of interesting consequences. First of all, due to a short non-equilibrium stage immediately after a merger, there are full loss cones and surges of tidal disruption events (TDE) \citep{2009ApJ...697L.149C}; this has been followed in more detail including stellar evolution and TDE \cite{2023ApJ...944..109L}. Furthermore, it has been suggested that the ``tidal'' disruption of hard binaries leads to hypervelocity stars escaping from the nucleus (and another member of the binary being tidally disrupted or absorbed by the SMBH, \citealp{2013ApJ...768..153Z}). In both cases, if compact objects are involved (neutron stars, stellar mass or intermediate-mass black holes) extreme or intermediate mass inspirals (EMRI, IMRI) with interesting gravitational wave emission signature will ensue \citep{2022MNRAS.516.1959M,2022ApJ...927L..18N}. In this paper, we do not discuss these effects, because we want to study the ``pure'' dynamics of the interaction of Kozai or other cycles of the triple SMBH with the stars. We do neither include stellar evolution or compact remnants, nor stellar binaries in our model. It is the subject of our future work - as a first step \citep{2023ApJ...944..109L} we published a study for TDE caused by different stars and used a method of scaling to extrapolate to the particle number of real galactic nuclei (which we cannot yet simulate directly). For EMRI and IMRI relativistic inspirals have already been included in the code \citep{2024MNRASinprep} EMRI and IMRI have received increasing also as possible sources for future space-based gravitational wave instruments \citep[cf. e.g.][]{2019PhRvD..99l3025A,2022MNRAS.510.2379V}.
}

\subsection{Stability of SMBH triples} \label{fourmodes}

\begin{table}
 \centering
  \begin{tabular}{lllll}
        \hline
Model   & Phase I	& Phase II	& Phase III	& Phase IV	\\
    \hline 
A     	& 16.50\%	& 	33.46\%		&	43.61\%	&	3.75\%		\\
B    	& 7.23\%	&  	60.79\%		&	23.11\%	&	2.36\%		\\
C		& 0.67\%	&	52.81\% 	&	21.75\% &	11.66\% 	\\    
D    	& 36.25\% 	& 	38.00\%		&	21.18\%	&	4.35\%		 \\
E       & 8.72\% 	& 	60.00\%		&	26.96\%	&	2.53\%		 \\
F  		& 27.30\%	& 	44.44\%		&	21.78\%	&	6.30\%		 \\

    \hline
  \end{tabular}
  \caption{Distribution of time fraction the SMBH triple systems in six simulations (Models A--F) spend on each phase (I=hierarchical; II=chaotic; III=binary+single phase; IV=unbound). This table does not include the occasional situations where $a_{\rm in}<0$ while $a_{\rm out}>0$. 
	\label{table:fourphase} }
\end{table} 

\begin{figure}
	\centering   
   	\includegraphics[width=0.45\textwidth,height=!]{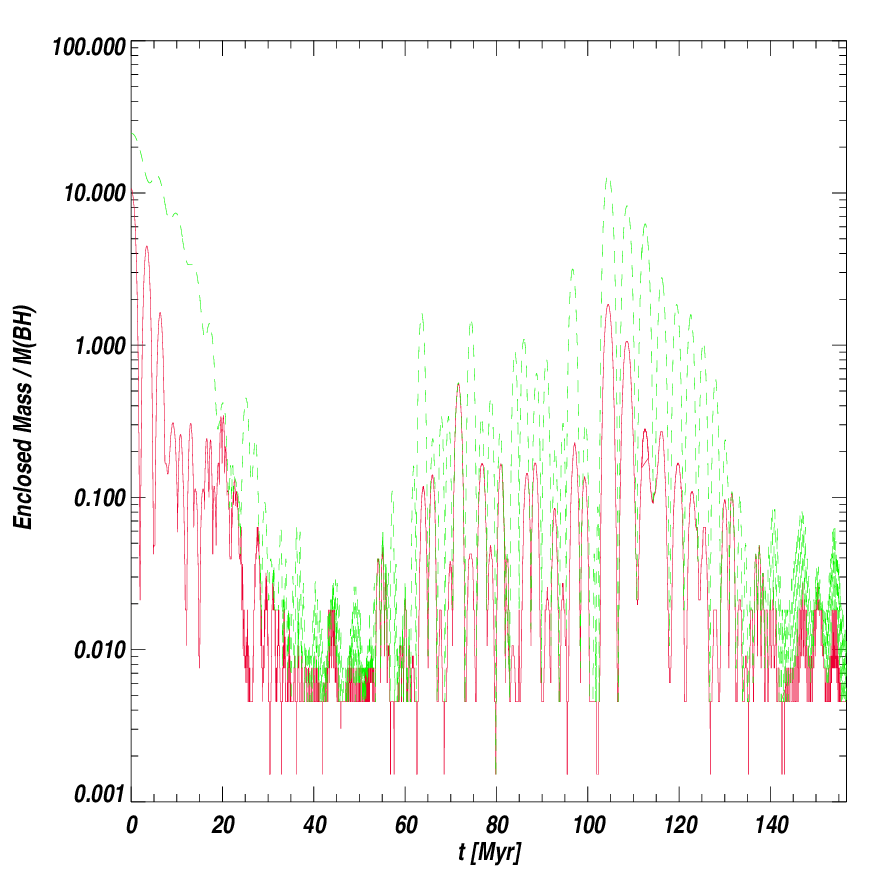}
   	\caption{The ratio between the enclosed stellar mass inside the orbital radius of the innermost SMB (red) and outermost SMBH (green) and the mass of a single SMBH as a function of time.
   	\label{figure:enclosedmass} }
\end{figure}


We define four phases to describe the dynamical status of the triple SMBH. To study the stability of triple systems, \cite{1995ApJ...455..640E} provided an empirical criterion by numerical experiments, in terms of a critical ratio $Y_{\rm min}$ of the periastron distance of the outer orbit to the apastron distance of the inner orbit. \cite{2001MNRAS.321..398M} improved the empirical criterion by using the resonance condition that the outer orbital angular velocity at periastron is equal to the inner orbit frequency. The early study on related topics can be traced back to the stability criteria study of Harrington \citep{1977AJ.....82..753H}. The implemented new criterion yields a  more stringent limit, especially for large $e_{\rm out}$ by making use of KS description. Thus, we computed the Mardling parameter $Y_{\rm m}$ as follows to examine the stability of the triple system 
\begin{equation}
	Y_{m} = C \left[ \left( {1 + \frac{m_3}{m_1 + m_2}} \right) 
	\frac{1 + e_{\rm out}}{ \left(1 - e_{\rm out} \right)^{1/2}} \right]^{2/5}  
	= \frac{p_{\rm crit}}{a_{\rm in}} 
	\quad,
\end{equation}
where $C = 2.8$ is an empirical constant, and $p_{\rm crit}$ is the critical value for the periastron separation of the outer binary. This equation is used to identify the escape of one body if the ratio between periastron separation of the outer binary $p_{\rm out}$ and semi-major axis of the inner binary $a_{\rm in}$ is smaller than the critical value given above by $Y_{m}$ (i.e. $ p_{\rm out}/a_{\rm in} > p_{\rm crit}/a_{\rm in}$). The equation above is only for coplanar prograde motion, and their later studies include an inclination fudge factor $Y_{\rm fac}$ for a wider application. Thus, we adopt 
\begin{equation}
	C = 2.8 Y_{\rm fac} = 2.8 \times (1-0.3i/180\degree)
\end{equation}
in this study. The result is shown in the upper-left panel of Figure~\ref{figure:keyparameters}. The figure shows that the system frequently switches between stable and unstable states, but tends to be generally unstable at the starting stage and become stable at the late stage. When the system is stable according to the criteria above, the triple SMBHs system can be considered as a stable hierarchical triple system. The binaries are considered unbound binaries if their total energy ($E_{\rm in}$ or $E_{\rm out}$) is positive. Using this modified Mardling criterion and the binding criterion, we classify the status of the triple SMBH system into four categories:
\begin{enumerate}
\item hierarchical merger phase  ($a_{\rm in}>0; a_{\rm out}>0$; and $0 < a_{\rm in}/p_{\rm out} < Y_{m}^{-1}$);
\item chaotic interacting phase  ($a_{\rm in}>0; a_{\rm out}>0$; and $a_{\rm in}/p_{\rm out} > Y_{m}^{-1}$);
\item binary plus single phase  ($a_{\rm in}>0$; and $a_{\rm out}<0$);
\item unbound phase ($a_{\rm in}<0$ and $a_{\rm out}<0$).
\end{enumerate}


For each simulation, we compute the fraction of time the SMBH triple system spends on each phase, and the results are shown in Table~\ref{table:fourphase}. Triples spend only a small fraction of time in the stable hierarchical merger phase, which is quite different from the results of isolated 3-body simulations shown in \S~\ref{section:comparison3body}.


We show the distance of the three SMBHs from the center-of-mass of the entire system as a function of time in Figure~\ref{figure:sixinone}. Although both the inner and outer binaries are unbound intermittently, the three SMBH black holes are generally bound. Also, we can find that violent encounters and exchanges of members between inner and outer binaries are common. Ejections are not observed.  

The center-of-mass of the stellar population (without SMBHs) is initially located at the coordinate origin. However, the center-of-mass of the entire system has a deviation after we add three SMBHs with different initial conditions. The  position of center-of-mass of the system, $\vec{r}_{\rm com}$, is defined as 
\begin{equation}
	\vec{r}_{com} = \frac{\sum{m_*\vec{ r_*}}+\sum{\mbh\rbh}}{\sum{m_*}+\sum{\mbh}}
	\quad, 
\end{equation}
and the velocity of center-of-mass $\vec{v_{\rm com}}$ 
\begin{equation}
	\vec{v}_{com} = \frac{\sum{m_* \vec{v_*}}+\sum{\mbh \vbh}}{\sum{m_*}+\sum{\mbh}}
	\quad,  
\end{equation}
where $m_{\rm *}$ represents the mass of each star and $m_{\rm BH}$ represents the mass of SMBHs, and the same for the position and velocity. Define the distance between the innermost SMBH and the position of center-of-mass to be $R_{\rm in}$, while the distance between the outermost SMBH and the position of center-of-mass to be $R_{\rm out}$. Assuming a sphere with a center at $\vec{r}_{\rm com}$ and a radius equal to $R_{\rm in}$/$R_{\rm out}$, then the total mass inside this sphere is defined as the enclosed stellar mass inside the orbital radius of the innermost/ outermost SMBH. 

In order to understand how the surrounding stellar population affects the dynamics of the triple SMBHs, we show the ratio between the enclosed stellar mass inside the orbital radius of the innermost/outermost  
SMBH and the mass of one SMBH in Figure~\ref{figure:enclosedmass}. At the start of simulation~D, the enclosed stellar mass for the innermost SMBH is approximately $10$ times the SMBH mass. 
The corresponding mass ratio for the outermost SMBH is roughly $25$. 
Both of these two quantities decrease dramatically in the first stage and the third stage and behave chaotically in the middle stage just as the evolution of their separation (see Section~\ref{section:orbitalevolution}). 
Most of the time, the mass of stars inside the orbital radius of the outermost SMBH is comparable to the mass of SMBH, showing that these stars together may play a comparable role as the outermost SMBH. This indicates that the stellar components of these galaxies can not be ignored when we study the dynamics of the SMBHs. 
For the region inside the magenta box of Figure~\ref{figure:fullevolution} where the example oscillation is located, the two ratios are rather low (around one percent of the SMBH mass) such that the steller components could not affect the triple effectively. This may be the prerequisite for the EKL mechanism to dominate the evolution of the triple SMBHs. 


\subsection{Periodic oscillations and their properties} \label{section:periodicoscillations}

\begin{figure}
	\centering   
   	\includegraphics[width=0.45\textwidth,height=!]{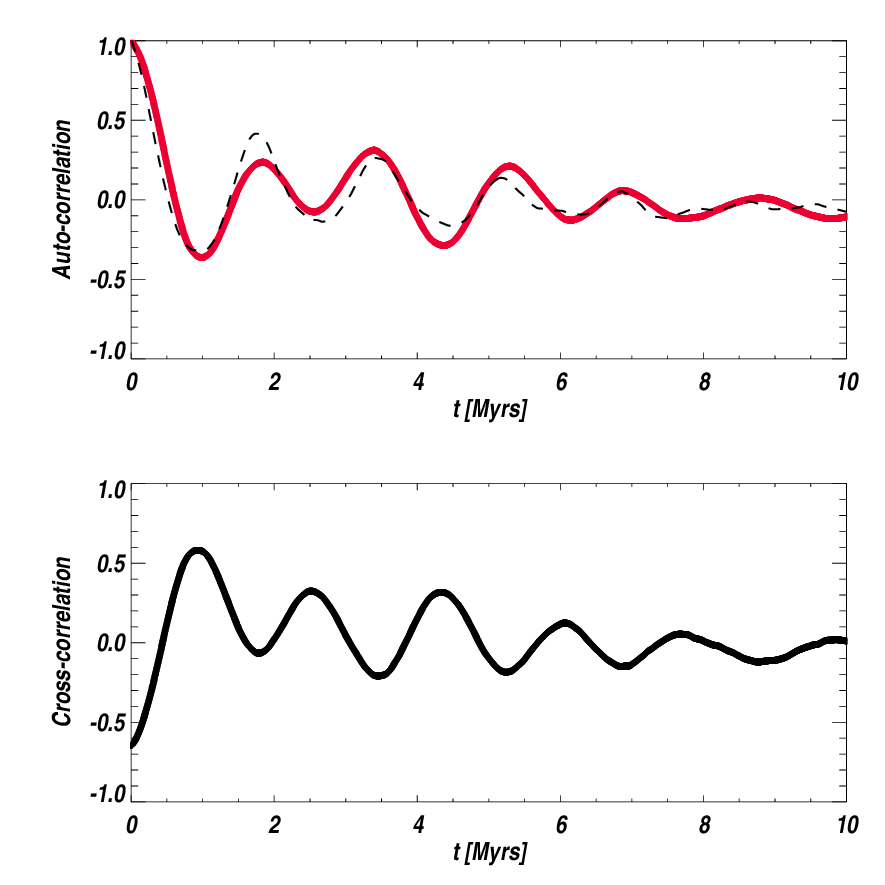}
   	\caption{Auto-correlation of inner eccentricity (red curve) and relative inclination angle (black curve) and cross-correlation between them. \label{figure:criterion} }
\end{figure}

\begin{figure*}
 	\vspace*{50pt}  
   	\includegraphics[width=\textwidth,height=!]{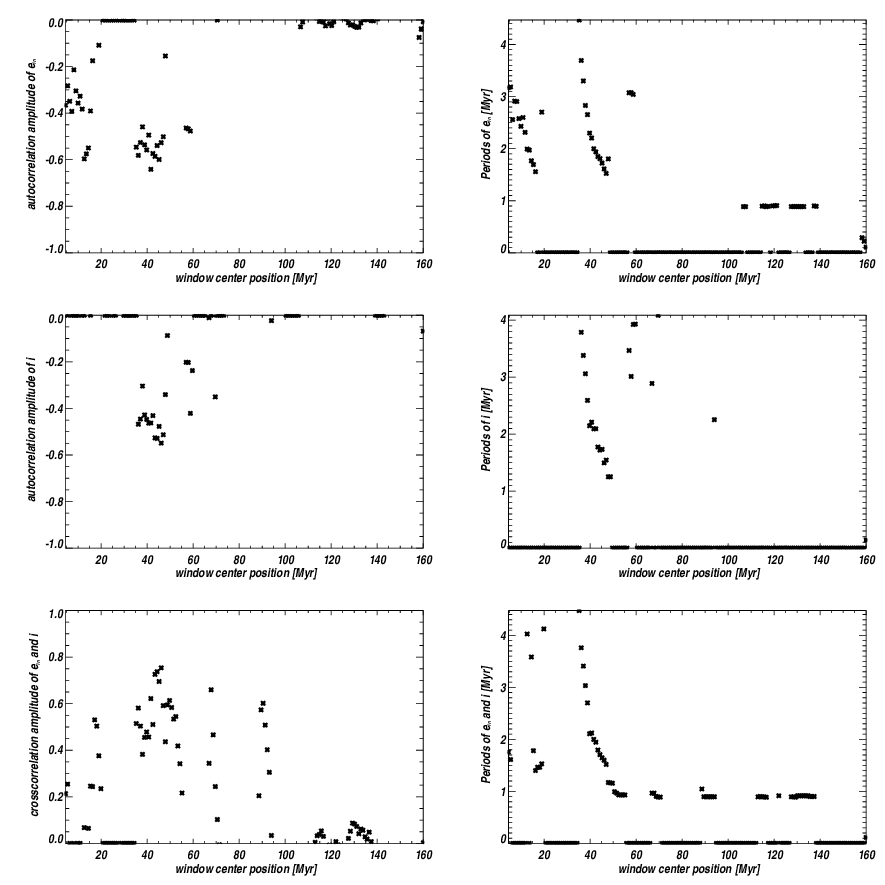}
   	\caption{Statistical results within different windows for auto-correlations and cross-correlations, as a function of time. {\em Upper-left panel:} the amplitude of the first trough in the correlation function curve of inner eccentricity within each window, oscillations are present once the amplitude is below zero (see the result of the example window in Figure~\ref{figure:criterion}). {\em Upper-right panel:} the value of the orbital periods at the first trough in the correlation function results. {\em Middle panel:} same as the top panel but for the mutual inclination angle. {\em Bottom panel:} same as the top panel but for the cross-correlation between the inner eccentricity and the mutual inclination, note that for cross-correlation we search for the first peak and the oscillations are present when the amplitude is above zero.
   	\label{figure:correlation} }
\end{figure*}

\begin{figure}
	\centering   
   	\includegraphics[width=0.5\textwidth,height=!]{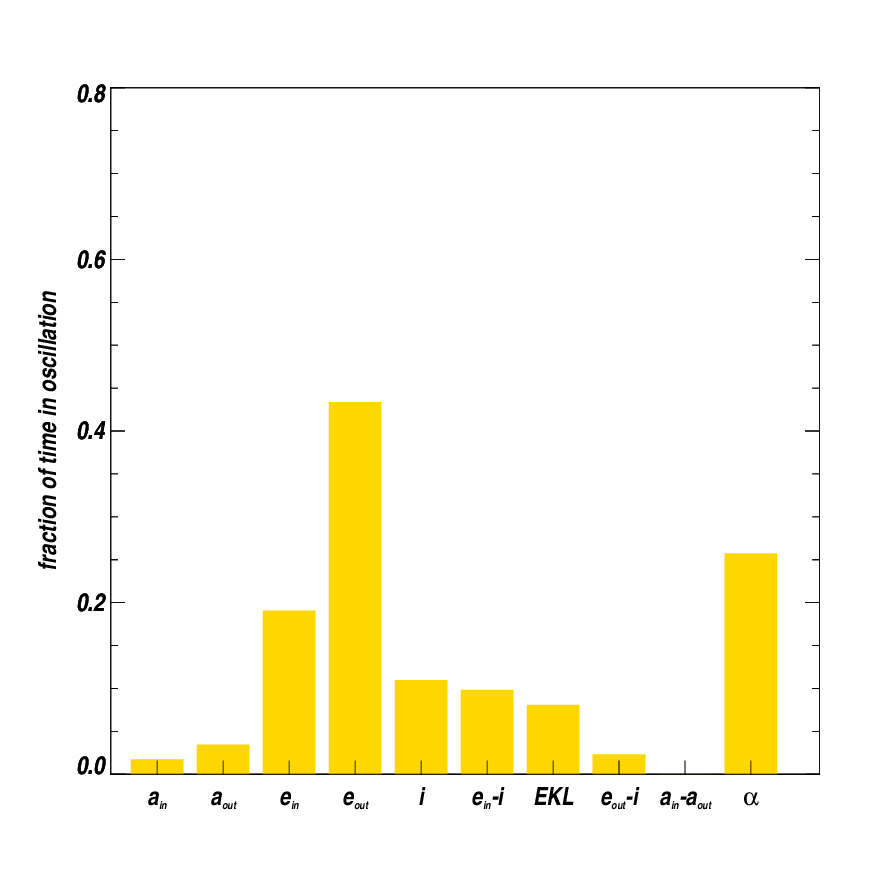}
   	\caption{Fraction of time that the orbital elements are in oscillation for model~D. The first five bars are the identification of auto-correlation results. The subsequent four bars are the results of the cross-correlation between the two parameters. The result labeled \lq{}EKL\rq{} represents the case in which the cross-correlation between $e_{\rm in}$ and $i$, and the auto-correlations of $e_{\rm in}$ and $i$ are all in oscillation. The quantity $\alpha$ at the right represents the fraction of time that either the inner binary or the outer binary is unbound.  
   	\label{figure:barplot} }
\end{figure}

\begin{figure}
	\centering   
   	\includegraphics[width=0.5\textwidth,height=!]{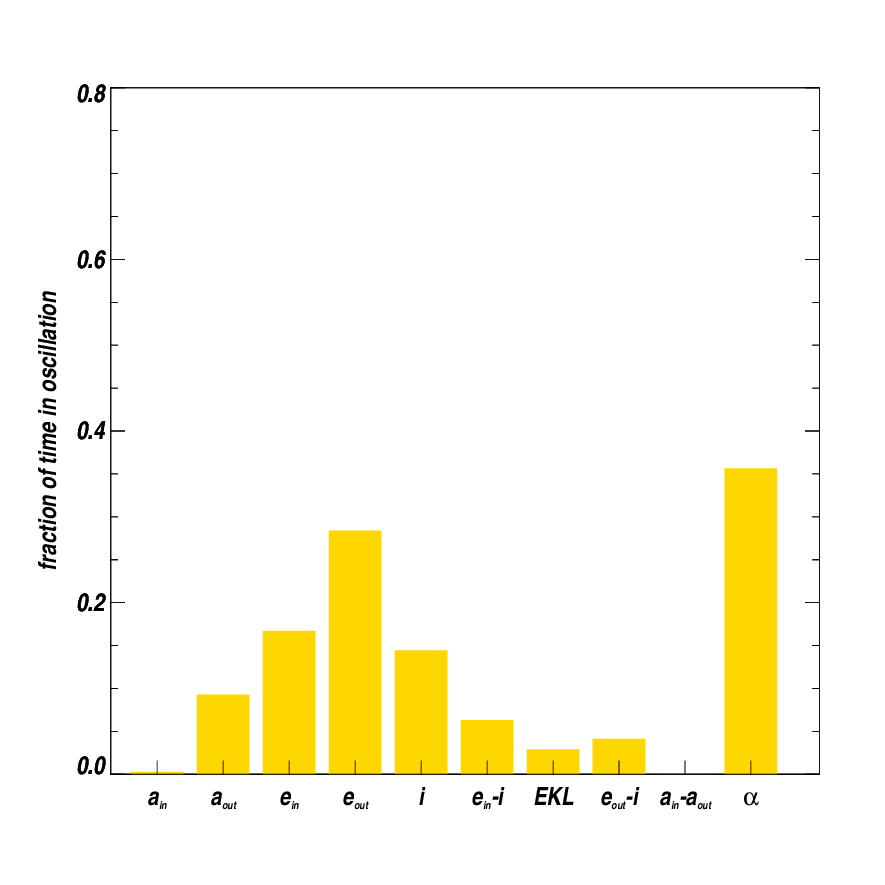}
   	\caption{Same quantities as in Figure~\ref{figure:barplot}, but this figure shows the averaged fraction of time for all six simulations. 
   	\label{figure:totalbarplot} }
\end{figure}


During the second and third stages, various \rosemary{periodic} oscillations are observed in the evolution of the orbital parameters between the inner and outer binary, which is the focus of this paper. These oscillations typically have a much longer period than the orbital periods of the inner and outer binary, so-called secular oscillations. An example of these secular oscillations is the Standard Kozai-Lidov oscillation as mentioned above. Under the SKL mechanism scheme, the $z$-component of the angular momentum of the inner binary is conserved, as a result, the eccentricity of the inner binary and the inclination angle oscillate together with a timescale $T_{\rm Kozai}$. \cite{2015MNRAS.452.3610A} found a numerical factor to better estimate this timescale based on previous studies:
\begin{equation}
	T_{\rm Kozai} \simeq \frac{8T_{\rm out}^2}{15\pi T_{\rm in}} \frac{m_1+m_2+m_3}{m_3}\left(1-e_{\rm out}^2\right)^{3/2}
	\quad,
  	\label{equation:tkozai} 
\end{equation}
where $T_{\rm in}$ and $T_{\rm out}$ denote the orbital period of inner and outer binary. 


After relaxing the test-particle approximation --- the high inclination ($>39.2\degree$) and the near-circular outer orbit assumption, the oscillation between the eccentricity and the inclination still occurs but becomes different from the SKL mechanism, especially its timescale. In this so-called EKL mechanism, the octuple-level approximation also plays an important role, and can even be stronger than the quadruple level. For the octuple level approximation, a similar estimation of oscillating timescale as $T_{\rm Kozai}$ is given by 
\cite{2011PhRvL.107r1101K} and \cite{2013MNRAS.431.2155N} as 
\begin{equation}
	T_{\rm oct} \sim \frac{T_{\rm Kozai}}{\epsilon_{\rm oct}}
	\quad,
\end{equation}
where $\epsilon_{\rm oct}$ represents the strength of the octupole level relative to the quadruple level of Newtonian Hamiltonian of hierarchical triples and is defined as follows:
\begin{equation}
	\epsilon_{\rm oct} = \frac{m_1-m_2}{m_1+m_2}\frac{a_1}{a_2}\frac{e_2}{1-e_2^2}
	\quad.
\end{equation}
%


As is discussed in previous sections, the motion of the equal mass triple SMBHs residing in the center of galaxy centers is complicated. In order to simplify this entangled problem, we leave the GR corrections and octuple level of EKL mechanism for future study. We study only the case of equal-mass SMBHs. Thus, $\epsilon_{\rm oct}=0$, and the octuple-level oscillations are not included. Also, these simulations do not include GR corrections, instead, we compute the GR timescale. The effect of GW emission may lead to an immediate merger of the inner black holes once the two SMBHs get close enough. Thus, it is necessary to look at the merger timescale as a result of gravitational wave emission. This time scale $T_{\rm gr}$ is defined by the ratio between semi-major axis $a_{\rm in}$, and secular evolution rate of semi-major axis ${\dot a_{in}}$:
\begin{equation}
	T_{\rm gr} = \frac{ a_{\rm in} }{ {\dot a_{\rm in}} } 
	\quad.
\end{equation}
The expression for the change in the semi-major axis ${\dot a_{\rm in}}$, is given by \cite{1964PhRv..136.1224P} 
\begin{equation}
	{\dot a_{\rm in}} = -\frac{64}{5}\frac{G^3 m_1 m_2 \left(m_1+m_2\right)}{c^5 a_{\rm in}^3 \left(1-e_{\rm in}^2\right)^{7/2}} \left(1+ \frac{73}{24} e_{\rm in}^2 + \frac{37}{96} e_{\rm in}^4 \right) 
	\quad,
\end{equation}
where $c$ is the speed of light. 
The \rosemary{adjusted} time scale $T_{\rm P}$ based on Peter\rq{}s formula given by \cite{2020MNRAS.495.2321Z} is expressed as follows: 
\begin{equation}
	T_{\rm P} =T_{\rm gr}8^{1-\sqrt{1-e_{\rm in}}}\exp(\frac{5G(m_1+m_2)}{c^2a_{\rm in}(1-e_{\rm in})})
	\quad.
\end{equation}
 The merger timescale required for the inner binary to coalesce only by emitting gravitational waves is shown in the bottom-right panel of Figure~\ref{figure:keyparameters}, for different moments in time.

The Kozai-Lidov precession rate $\dot \omega_{\rm KL}$ compared to the GR precession rate $\dot \omega_{\rm GR}$ can be expressed as \citep{2011ApJ...729...13C, 2016MNRAS.461.4419B}
\begin{equation}
	\frac{\dot \omega_{\rm KL}}{\dot \omega_{\rm GR}} \simeq \frac{c^2m_3a_{\rm in}^4(1-e_{\rm in}^2)^{1/2}} {G\left(m_1+m_2\right)^2a_{\rm out}^3\left(1-e_{\rm out}^2\right)^{3/2}}
	\quad.
  	\label{equationprecession} 
\end{equation}


To search for such periodic oscillations of a dynamical system over a certain time interval in the simulation data, frequency analysis is a powerful technique. Unfortunately, the common method of Fourier transformation analysis is not suitable to study the periodic properties of those orbital parameters discussed above, since oscillations here are relatively short-lived (typically less than ten periodic cycles) and with dramatic attenuation on both amplitude and period of the oscillation. In this situation, the results obtained from Fourier analysis are usually too weak. Thus, we use the auto-correlation function of an orbital element or cross-correlation between two orbital elements as an approach to identify the oscillations and determine their periods.  


We illustrate our method of investigating the periodic oscillations in Figure~\ref{figure:criterion} by showing an example of the oscillating region inside of the magenta box in Figure~\ref{figure:fullevolution}. The eccentricity of the inner binary $e_{\rm in}$ and relative inclination angle $i$ are both in clear oscillations within the region inside the magenta box. This is a secular oscillation according to the orbital timescale of the inner orbit shown in Figure~\ref{figure:keyparameters} (upper-right panel).
In Figure~\ref{figure:criterion}, the upper panel shows the amplitude of auto-correlations of $e_{\rm in}$ (red) and $i$ (black). The lower panel shows the cross-correlation between $e_{\rm in}$ and $i$. The oscillation for each quantity is clear, but their strength (represented by the amplitude) is decreasing irregularly and dramatically. The amplitude of the first minimum of the auto-correlation of both $e_{\rm in}$ and $i$ damped to a value around $-0.4$, while the first maximum of the cross-correlation damped to about $0.6$. Note that the cross-correlation result shows $e_{\rm in}$ and $i$ are in anti-phase. The period of the oscillation also decreases over time. During this process, the eccentricity of the inner binary is effectively excited (at the peaks of this oscillation $e_{\rm in}$ is close to unity, see Figure~\ref{figure:fullevolution}). When the orbits of the inner binary or the outer binary temporarily become parabolic or hyperbolic, the semi-major axis cannot be used. But their eccentricity is valid and could still depict the evolution of their angular momentum. The fraction of time that either the inner binary or the outer binary is unbound is shown in the right-most of Figure~\ref{figure:barplot}. 


Our paper aims to find secular oscillations with periods much longer than their orbital periods. From the upper-right panel of Figure~\ref{figure:keyparameters} we find that the orbital period of the inner binary ranges from hundreds of years to 0.1~Myr. Then the corresponding secular oscillations should be at least tens or hundreds times this orbital period. On the other hand, from the bottom-left panel of Figure~\ref{figure:keyparameters}, we find that the typical timescale for EKL cycles is from 0.5~Myr to 10~Myr (ignoring the sharp spikes) when the triple is in a hierarchical merger phase. Also, the window size can not be too large since these secular oscillations are typically short-lived. We, therefore, select a window size of about 10~Myr to cut the data and use the correlation function to analyze the oscillation inside each window. The result of an example window(the magenta box in Figure~\ref{figure:fullevolution}) is shown in Figure~\ref{figure:criterion}. Then we use this window to go over the whole data, with a step size equal to one-tenth of the window size, to avoid the missing discoveries caused by the selection of the window position. 


The result of the correlation analysis on auto-correlation of $e_{\rm in}$ and $i$ is shown in the upper and middle panels of Figure~\ref{figure:correlation}, and the results for cross-correlation between $e_{\rm in}$ and $i$ are shown in the lower panel. The left three plots show the amplitude at the first minimum of the auto-correlation function (top and middle panel) or the amplitude at the first maximum of the cross-correlation function (lower panel). Periods of the oscillations are plotted respectively on the right. For most identified oscillations in $e_{\rm in}$ and $i$, their amplitudes at the first peak or trough spread from 0 to 0.8, while the identified periods spread from 1~Myr to 4~Myr. 


In the region corresponding to the magenta box, the periods of $e_{\rm in}$ and $i$ are from about 4~Myr down to 0.8~Myr. The decreasing orbital periods may be attributed to either the irregular evolution of EKL mechanism of the triple as the example shown in Figure~\ref{figure:3bodyexample23.6} or the perturbation of the surrounding stellar population. While the typical timescale computed from Equation~(\ref{equation:tkozai}) is from about 0.6~Myr down to 0.1~Myr, which can also be found in the down-left panel of Figure~\ref{figure:keyparameters}. 
If we define the hardening rate as 
\begin{equation}
	s = \frac{d}{dt} \left(\frac{1}{a_{\rm in}}\right) 
	\quad, 
\end{equation}
then the average hardening rate within the magenta box is $0.009$~pc$^{-1}$Myr$^{-1}$, which is even lower than the value of the average hardening rate over the whole simulation $0.016$~pc$^{-1}$Myr$^{-1}$. This result indicates that when we simulate without including general relativity effects, the EKL mechanism could not offer a dominating impact on accelerating the merger process of the inner binary in this example region, even the eccentricity is excited to the value very close to unity repeatedly. This agrees with our current understanding that EKL mechanism itself could not contribute to orbital shrinkage of the inner SMBHB without GR and tides, no matter whether we have a surrounding cluster or not. Here we record the value of the average hardening rate since it would be interesting to compare this result with simulations including GR in future studies. Also, when we compute the averaged hardening rate for all identified EKL oscillations and the averaged hardening rate for all six simulations, we get similar results as the example discussed above. 


Given the fact that even for the one with the best example with clear oscillation, the amplitude of the first minimum/maximum is weak due to the irregular shape of the oscillation curve and their decreasing orbital periods. Thus, we set the critical value of the power spectrum of an identified oscillation at the first trough to be 0.1, which is a relatively low standard, in order to include those irregular oscillations. For oscillations with stronger amplitude at the first minimum than this critical value, we count them as identified oscillations. Then based on the number of these identified oscillations in each window, we plot the statistical results in Figure~\ref{figure:barplot}. 

\hw{ If we use a higher value as the critical value for the power spectrum of identified oscillation, for example, 0.5, then the result can be quite different. The total number of identified oscillations would be less than half of the current value. The strength of the oscillation for both eccentricity and inclination angle reaches their largest value at around 0.5 for eccentricity and 50 degrees for inclination. Oscillations with very small eccentricity or inclination are rare since the triples are unstable most of the time. Both eccentricity and inclination angle stay with relatively high values except in the late stage when the inner binary is hard enough shortly before they start the merging process. }

Few oscillations are found in $a_{\rm in}$ and $a_{\rm out}$ according to Figure~\ref{figure:barplot}, and one of the reasons for this is that when the binaries are unbound the semi-major axis is not defined for hyperbolic orbits. The fraction of time that $e_{\rm in}$, $e_{\rm out}$, and $i$ are in oscillation is relatively high. The fraction of time that $e_{\rm in}$ and $i$ are correlated is only slightly smaller than the fraction of time that $i$ is in oscillation, showing a good covariation between these two quantities. This is different from the situation of the eccentricity of the outer binary $e_{\rm out}$ - it has the most fraction of time in oscillation but its cross-correlation with $i$ is lower. When the eccentricity of the inner binary $e_{\rm in}$, the inclination angle $i$ and the cross-correlation between $e_{\rm in}$ and $i$ are all in oscillation at the same time, we refer it as the identified EKL oscillation. In this example simulation, the SMBH triple is under EKL oscillation for about 8\% of the time. The quantity with the most fraction of time in oscillation is  $e_{\rm out}$ (43\%), and even much greater than the number of oscillations in $e_{\rm in}$ (19\%) and $i$ (11\%). Unlike for $e_{\rm in}$, the link between $e_{\rm out}$ and $i$ is relatively weak, but not negligible. 


Also, we plot the averaged results for all six simulations similar to Figure~\ref{figure:barplot} in Figure~\ref{figure:totalbarplot}. The average fraction of time that the triples are in EKL oscillation is only 3\%, which is relatively low. Mean motion resonances (MMR) are found and discussed mainly in planetary systems with high mass ratios to the central body. But when the ratio of the two orbital periods between our inner binary and outer binary are occasionally close to the ratio of two small integers, MMR may also occur. Thus, we try to find these MMR-like oscillations from our simulation data as well. Unfortunately, as is shown in Figures~\ref{figure:barplot} and~\ref{figure:totalbarplot}, the cross-correlation results between $a_{\rm in}$ and $a_{\rm out}$ show that no obvious MMR is observed, indicating the chance for forming this type of resonance is small. 
 

\section{Comparison three-body simulations} \label{section:comparison3body}

\begin{figure}
	\centering   
   	\includegraphics[width=0.5\textwidth,height=!]{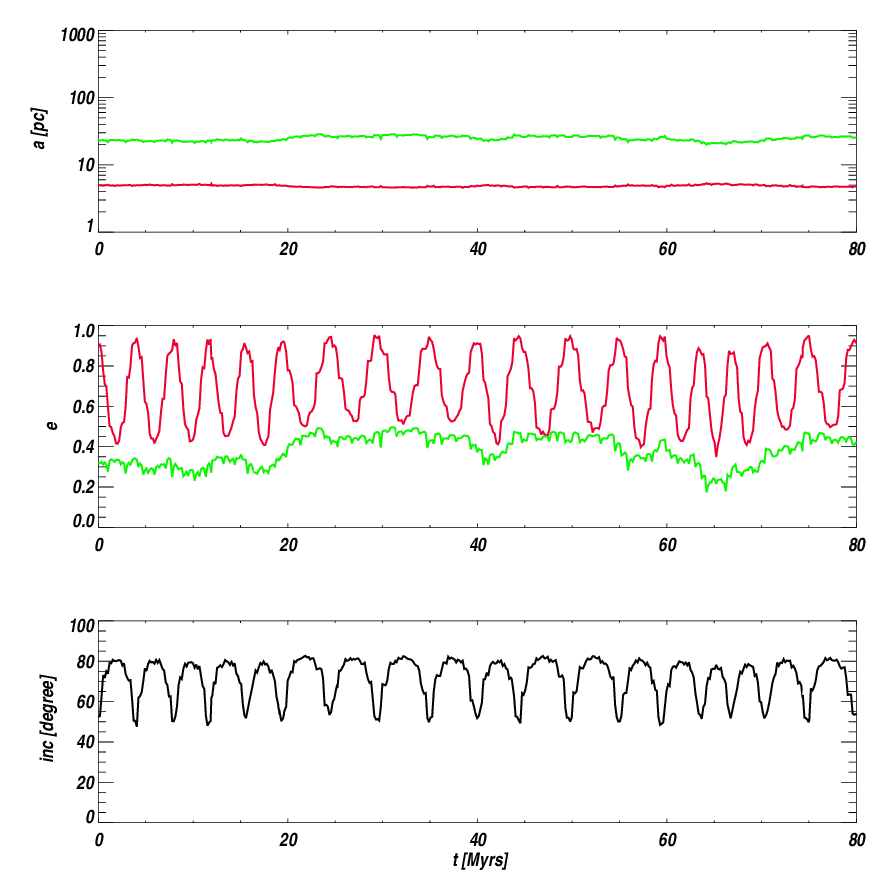}
   	\caption{Evolution of the orbital elements of an isolated hierarchical triple SMBH within the first $t=80$~Myr. The initial conditions for three SMBHs are obtained from the snapshot of model~D at $t=36.8$~Myr at the beginning of the magenta box. \tk{Unlike the triple in the clustered environment (shown in Figure~\ref{figure:fullevolution}), the EKL oscillation in this isolated system persists much longer than $15$~Myr, and the distortion of the oscillation curve is limited.}
   	\label{figure:3bodyexample} }
\end{figure}

\begin{figure}
	\centering   
   	\includegraphics[width=0.5\textwidth,height=!]{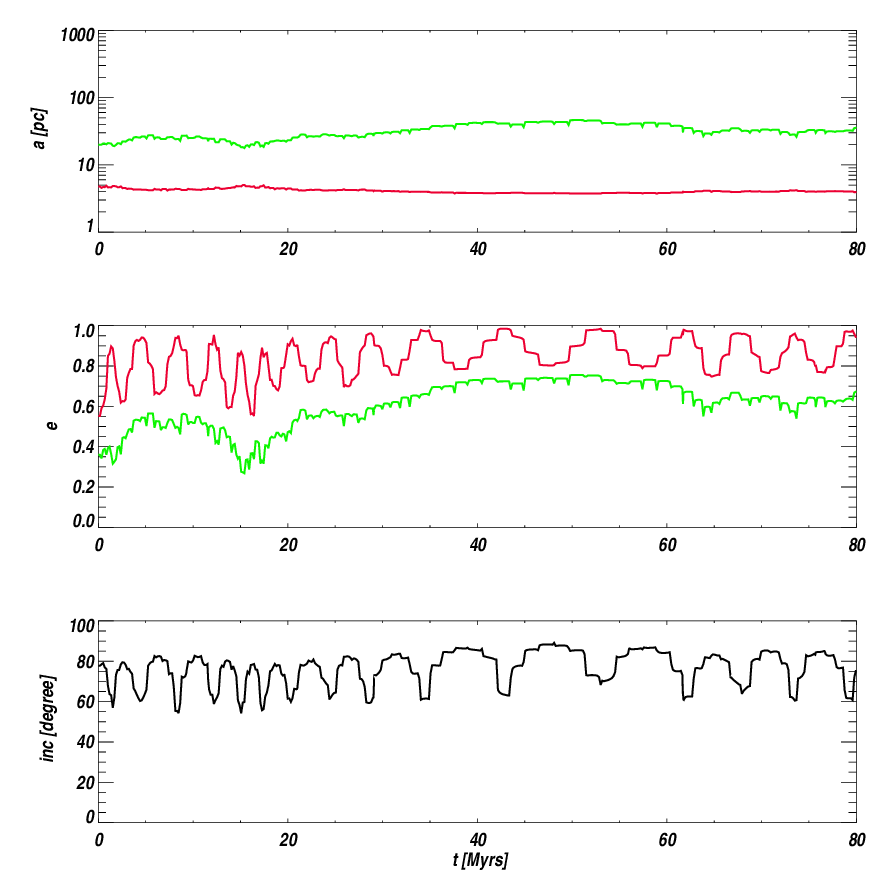}
   	\caption{Evolution of the orbital elements of an isolated hierarchical triple SMBH. The initial conditions for three SMBHs are obtained from the snapshot of model~D at $t=40.0$~Myr. 
   	\label{figure:3bodyexample23.6} }
\end{figure}


To verify if the identified oscillations above can be reproduced by pure three-body interactions, especially three-body interaction under EKL mechanism, we carry out simulations with the SMBH triples only with the same initial conditions as the above $N$-body simulation. We take $20$ snapshots with intervals of $0.17$~Myr, starting from $t=36.8$~Myr in the part of the simulation shown in the magenta box of Figure~\ref{figure:fullevolution}, and extract the kinematical data for the three SMBHs. Then we carry out three-body simulations using the \texttt{CHAIN} code \citep{1993CeMDA..57..439M, 2003gnbs.book.....A} to study the evolution of the isolated three SMBHs for a timescale of $80$~Myr. With a focus on studying the effect of surrounding stellar components rather than the effect of general relativity --- which has been studied by \cite{2016MNRAS.461.4419B}, we use CHAIN instead of ARCHAIN in this study. 


As an illustration, we show the first simulation at $t=36.8$~Myr out of the $20$ simulations in Figure~\ref{figure:3bodyexample}, and the last simulation at $t=40.0$~Myr in Figure~\ref{figure:3bodyexample23.6}. The SMBH triples clearly follow the  EKL mechanism --- the inner eccentricity and inclination oscillating with the same period at the anti-phase. Note that we have equal-mass triple SMBHs, so the oscillation would not be as perfect as those in SKL, and do not include the octuple-level oscillation. We successfully reproduced the identified EKL oscillation, but they are different from the original oscillation with surrounding stellar populations. For example, at time $t=36.8$~Myr, the oscillation period obtained from Figure~\ref{figure:criterion} is about $1.8$~Myr, while that of three body simulations obtained from Figure~\ref{figure:3bodyexample} is about $4.5$~Myr. The averaged oscillation period we obtain from these twenty simulations is $4.8$~Myr. This is slightly longer but generally agrees with the oscillation periods get from Figure~\ref{figure:correlation} when $t=36.8 - 40.0$~Myr, which are about $4$~Myr down to $2$~Myr. Note that the oscillation in three body simulations lasts much longer than the original oscillation shown in the magenta box of Figure~\ref{figure:fullevolution}. 


Also, the periods vary dramatically or decrease with time instead of keeping the same as in SKL oscillations as is shown in Section~\ref{section:periodicoscillations}, especially in Figure~\ref{figure:3bodyexample23.6}. The reason for this deviation may be attributed to the mildly hierarchical configuration of our SMBH triples with equal mass. Within each three-body simulation, if the maximum period is more than twice the minimum period, we define this set of simulations as an irregular oscillation. The number of irregular oscillations, such as the example shown in Figure~\ref{figure:3bodyexample23.6}, is ten out of twenty in total. Some of these irregular oscillations even break down within $80$~Myr, and some of them break down in the next hundred Myr. This suggests that, although these three-body systems are in EKL oscillation, they are already on an irregular oscillation and even close to breaking down. 
\tk{The value of $p_{\rm out}/a_{\rm in}/Y_{\rm m}$ in the upper left panel of Figure~\ref{figure:keyparameters} shows that the triple system is very close to, but below the stability boundary at $t=36.8-40.0$~Myr. The system is therefore unstable, with a long instability timescale. This can also be verified by the variations in $a_{\rm in}$ in Figure~\ref{figure:3bodyexample}. If the system were stable, $a_{\rm in}$ would be constant, apart from the slight variations on the orbital period timescale. Nevertheless, when the oscillation starts at $t=36.8$~Myr, the isolated triple is relatively more stable than at the end stage, at around $t=40.0$~Myr. This is further supported by the decrease of $p_{\rm out}/a_{\rm in}/Y_{\rm m}$ in Figure~\ref{figure:keyparameters}. When in isolation, the SMBH triple could evolve more than 10 Gyr before their EKL oscillation breaks down, and the amplitude and the duration of each period of the oscillation evolve dramatically slower than the one in the magenta box.
}

Note that not every identified oscillation in the $N$-body simulation can be produced by the three-body simulation. In fact, even the best examples of identified oscillation besides the one in the magenta box could hardly be repeated with the three-body simulation using the same initial condition. While the other particles besides SMBHs are removed, the triple SMBHs are sometimes unbound to each other, or bound but not in oscillation, or in oscillation but on a very different timescale (e.g., on a much longer timescale; up to several Gyr). 


\section{Discussion and Conclusions} \label{section:conclusions}


\hw{We have analyzed the evolution of several pathfinding numerical simulations of triple SMBHs in the center of surrounding star clusters with a mass one hundred times each SMBH mass.} Depending on the stability status of SMBH triples, we have defined four characteristic phases and summarised the time fraction the triple systems spend in each phase. 


We find that unlike pure 3-body simulations shown in Section~\ref{section:comparison3body}, the triples spend only a small fraction of their time in the hierarchical merger phase, indicating that the surrounding stellar population plays an important role. This is supported by the evidence that most of the time, the enclosed stellar mass within the orbits of the innermost or the outermost SMBH is comparable to the mass of SMBH. In the example simulation~D, at a region inside the magenta box where the enclosed stellar mass is relatively low, an oscillation similar to the EKL oscillation is observed. Although it is one of the long-lived examples observed, it lasts for only several periods, with its amplitude damping and its orbital period shrinking with time. We further confirmed this oscillation to be a disturbed EKL oscillation in two ways. First, we use the correlation function analysis to confirm that (i) $e_{\rm in}$ and $i$ are both in oscillation, (ii) the oscillation found in $e_{\rm in}$ and $i$ are correlated with an anti-phase or in-phase and (iii) the oscillation period is comparable to the typical timescale of EKL oscillation. 


We extract the dynamical information of the SMBH triples and use this information as the initial conditions for the comparison of three-body simulations. The averaged oscillation period obtained from these simulations generally agrees with the oscillation periods obtained by the correlation function. We also found that for these equal-mass SMBH triples, EKL oscillations are distorted and even become irregular. We use the correlation function to find oscillations over the entire data and show the statistical results for each type of oscillation. 


The overall average fraction of time that the triples are in EKL oscillation for all six simulations is only 3\%. In simulation~D, the SMBH triple is under EKL oscillation for about 8\% of the time. The averaged hardening rate of the inner binary during the EKL oscillations is $0.009$~pc$^{-1}$Myr$^{-1}$, which is even lower than the averaged hardening rate within the full-time simulation. This suggests that the role of EKL oscillations in accelerating the merger process of the inner binary may have been overestimated in previous studies. We also find that the orbital parameter with the largest fraction of time in oscillation is  $e_{\rm out}$, and the link between $e_{\rm out}$ and {i} is relatively weak, but not negligible. No MMR-like oscillation is observed, indicating the chance of forming this kind of resonance is rather low. 

Several physical processes have not been included in our current study, which may be considered in future investigations. Our study is limited to major mergers with equal-mass SMBHs, and our future study on unequal-mass systems may be fruitful. The gravitational interaction of the triple SMBH system is coupled to other processes such as the mass growth of the SMBH, tidal effects, and gravitational wave emission due to the spin of the SMBH, etc. Similarly, \cite{1998AJ....116..444M} studied the mass transfer in the secular evolution of the triple system CH~Cygni, triggered by eccentricity oscillation of the inner binary. Also, tidal interaction together with mass transfer for compact x-ray binaries in hierarchical triples has been studied \citep{1987ApJ...317..737B}. 


\section*{Acknowledgments}
\tk{We express our gratitude to the anonymous referee, whose comments and suggestions helped to improve this paper.}
W.H. and R.S. thank Thorsten Naab for helpful discussions, suggestions, and support. M.B.N.K. acknowledges support from the National Natural Science Foundation of China (NSFC, grant 11573004). This research was supported by the Research Development Fund (grant RDF-16-01-16) of Xi'an Jiaotong-Liverpool University (XJTLU). R.S. thanks M.B.N.K. and the astronomy team in Suzhou for their hospitality during several visits to XJTLU, during which part of this work was done. R.S. acknowledges the support of NSFC grants 11721303, 11988101 and 11303039, of the Key International Partnership Program of the Chinese Academy of Sciences (CAS) (No.114A11KYSB20170015), of the Strategic Priority Research Program (Pilot B) ``Multiwavelength gravitational wave universe'' of Chinese Academy of Sciences (No.XDB23040100), and of Yunnan Academician Workstation of Wang Jingxiu (No. 202005AF150025). The authors gratefully acknowledge the Gauss Centre for Supercomputing e.V. for funding this project by providing part of the computing time used in this project through the John von Neumann Institute for Computing (NIC) on the GCS Supercomputer JUWELS at J\"ulich Supercomputing Centre (JSC). As computing resources, we also acknowledge the Silk Road Project GPU systems and support by the computing and network department of NAOC. We acknowledge the funds from the ``European Union NextGenerationEU/PRTR'', Programa de Planes Complementarios I+D+I (ref. ASFAE/2022/014).


\section*{Data Availability}
The data underlying this article will be shared on reasonable request to the corresponding author. 

\bibliographystyle{mn2e}
\bibliography{./references}


\appendix 

\section{Additional results} \label{section:appendix}

\begin{figure*}
 	\vspace*{50pt}  
   	\includegraphics[width=\textwidth,height=!]{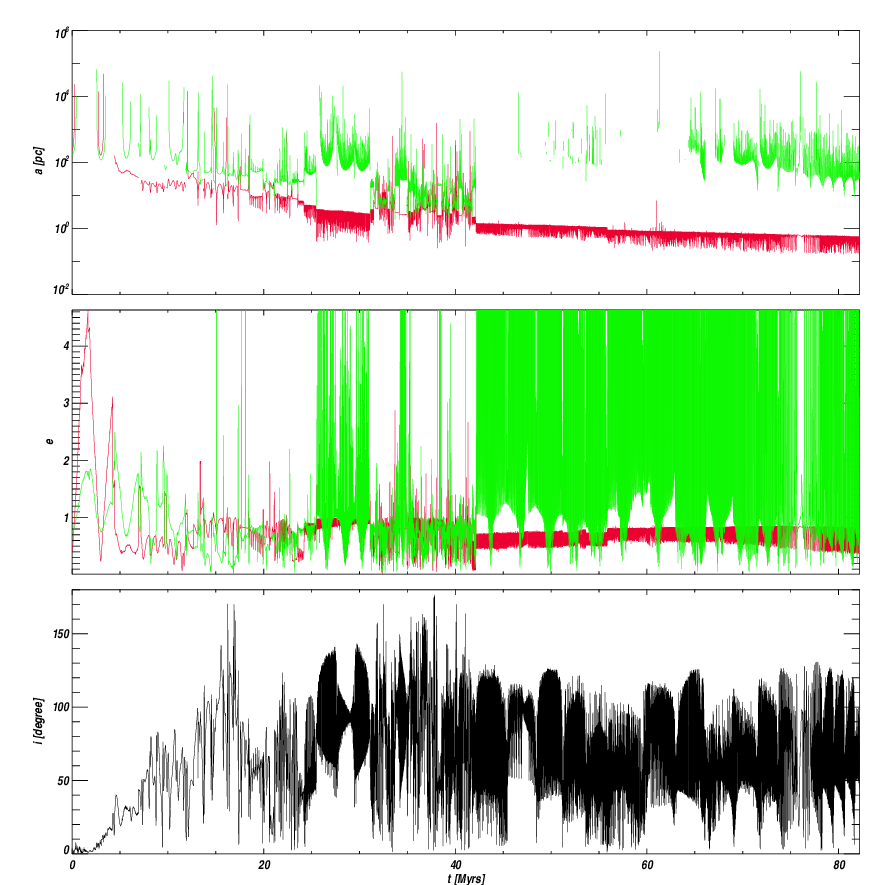}
   	\caption{Evolution of the semi-major axis, eccentricity, and relative inclination in simulation~A.
   	\label{figure:fullevolution1} }
\end{figure*}

\begin{figure*}
 	\vspace*{50pt}  
   	\includegraphics[width=\textwidth,height=!]{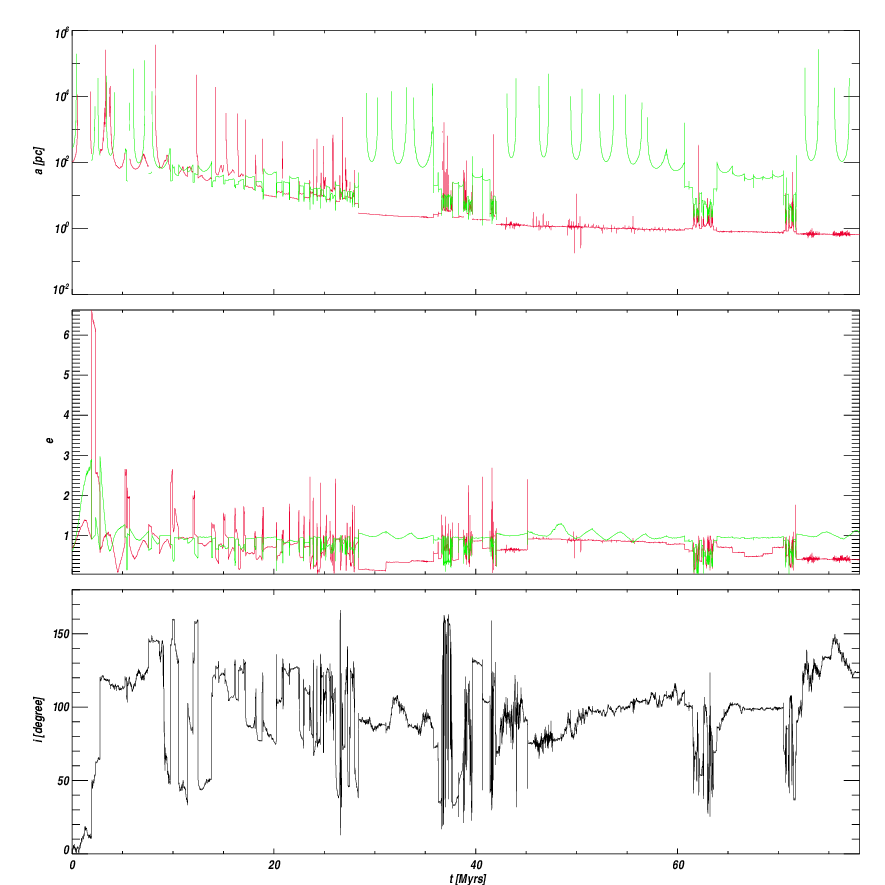}
   	\caption{Evolution of the semi-major axis, eccentricity, and relative inclination in simulation~B. 
   	\label{figure:fullevolution2} }
\end{figure*}

\begin{figure*}
 	\vspace*{50pt}  
   	\includegraphics[width=\textwidth,height=!]{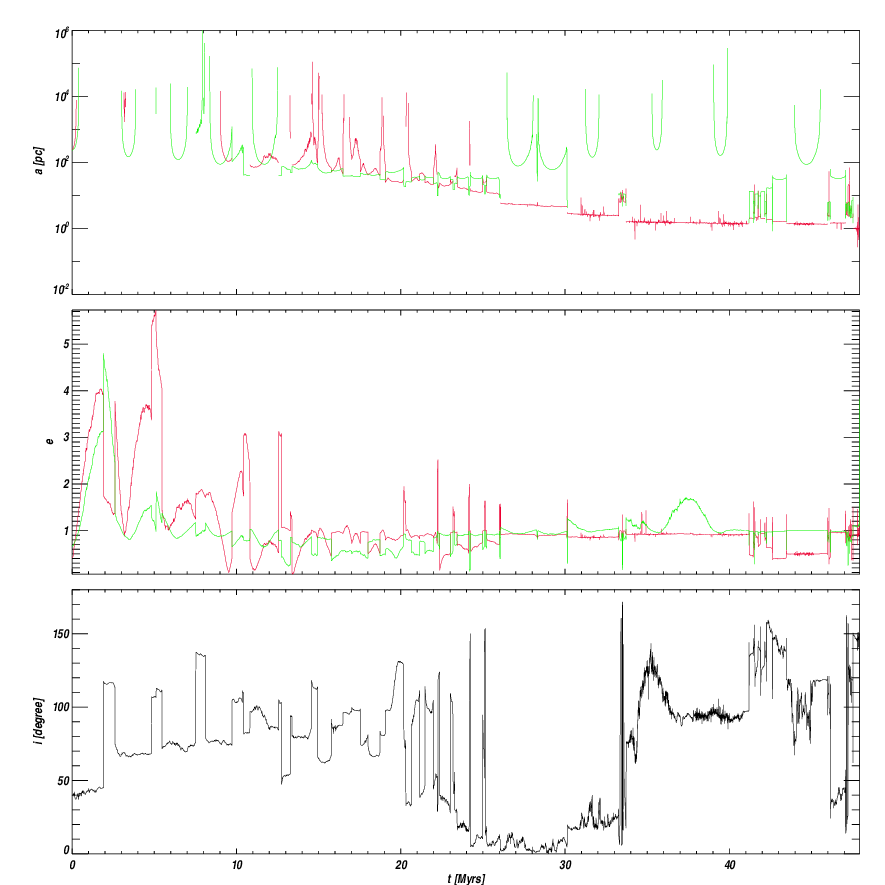}
   	\caption{Evolution of the semi-major axis, eccentricity, and relative inclination in simulation~C. 
   	\label{Figure:fullevolution3} }
\end{figure*}

\begin{figure*}
 	\vspace*{50pt}  
   	\includegraphics[width=\textwidth,height=!]{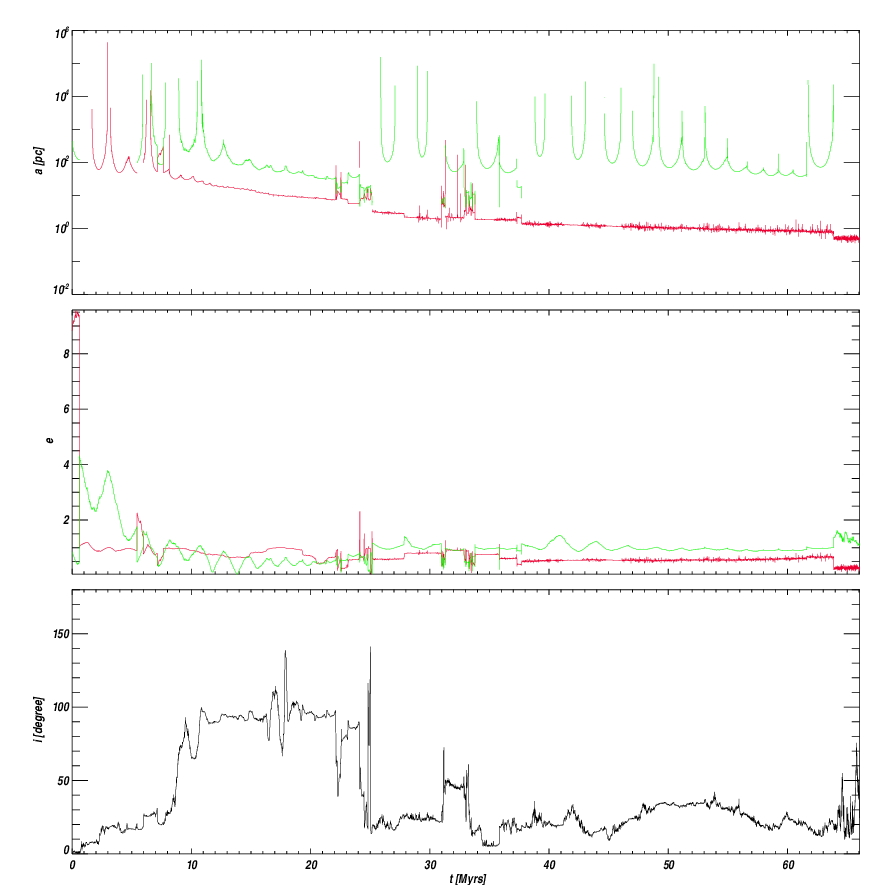}
   	\caption{Evolution of the semi-major axis, eccentricity, and relative inclination in simulation~E. 
   	\label{figure:fullevolution4} }
\end{figure*}

\begin{figure*}
 	\vspace*{50pt}  
   	\includegraphics[width=\textwidth,height=!]{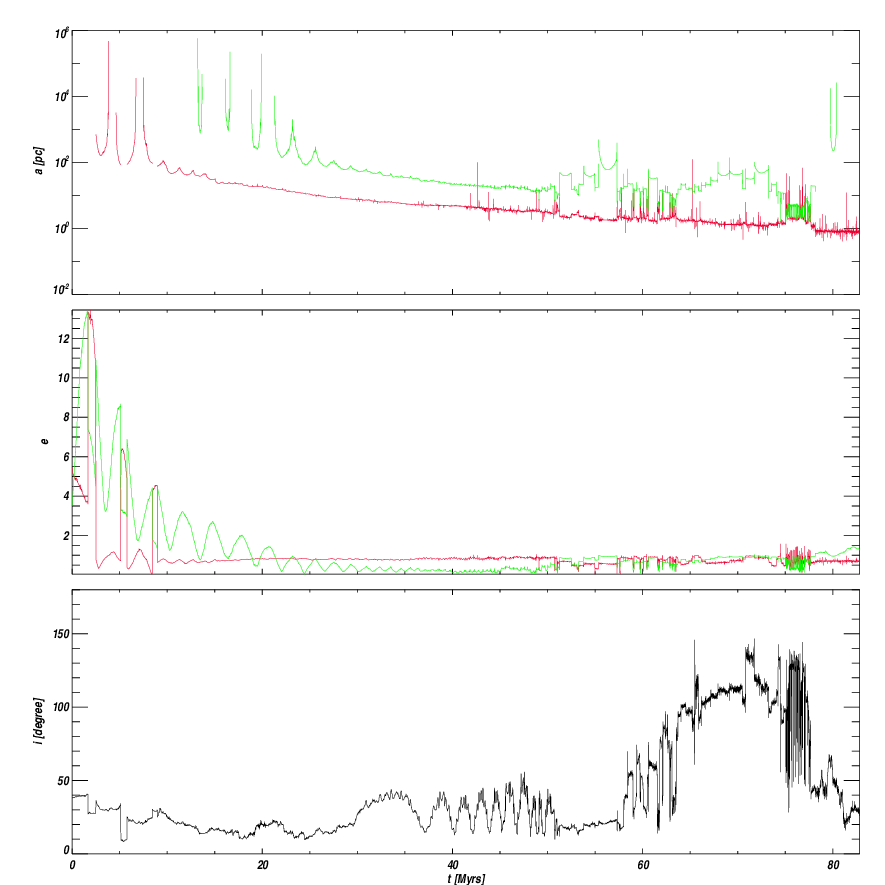}
   	\caption{Evolution of the semi-major axis, eccentricity, and relative inclination in simulation~F.
   	\label{figure:fullevolution5} }
\end{figure*}

Detailed evolution results for models A, B, C, E, and F are shown in Figures~\ref{figure:fullevolution1}-\ref{figure:fullevolution5}. The results for model~D are shown in Figure~\ref{figure:fullevolution}.

\end{document}